\documentclass[a4paper,11pt]{article}
\usepackage{amssymb, amsmath}
\usepackage[curve]{xypic}

\providecommand{\norm}[1]{\lVert#1\rVert}

\providecommand{\Ker}{\textnormal{Ker}}
\providecommand{\IIm}{\textnormal{Im}}

\providecommand{\Hom}{\textnormal{Hom}}

\providecommand{\Ker}{\textnormal{Ker}}

\providecommand{\Int}{\textnormal{Int}}

\providecommand{\PD}{\textnormal{PD}}
\providecommand{\cpt}{\textnormal{cpt}}

\providecommand{\Maps}{\textnormal{Maps}}

\providecommand{\vol}{\textnormal{vol}}

\catcode`\@=11

\def\section{\@startsection{section}{1}{\z@}{3.5ex plus 1ex minus
   .2ex}{2.3ex plus .2ex}{\LARGE\bf}}
\def\subsection{\@startsection{subsection}{1}{\z@}{3.5ex plus 1ex minus
   .2ex}{2.3ex plus .2ex}{\Large\bf}}
\def\subsubsection{\@startsection{subsubsection}{1}{\z@}{3.5ex plus 1ex minus
   .2ex}{2.3ex plus .2ex}{\large\bf}}

\begin{document}

\begin{titlepage}
\titlepage
\rightline{SISSA 52/2009/FM}
\vskip 2.5cm
\centerline{ \bf \huge Topics on the geometry of D-brane}
\vskip 0.5cm
\centerline{ \bf \huge charges and Ramond-Ramond fields}
\vskip 2truecm

\begin{center}
{\bf \Large Fabio Ferrari Ruffino}
\vskip 1.5cm
\em 
International School for Advanced Studies (SISSA/ISAS) \\ 
Via Beirut 2, I-34151, Trieste, Italy\\
and Istituto Nazionale di Fisica Nucleare (INFN), sezione di Trieste

\vskip 2.5cm

\large \bf Abstract
\end{center}

\normalsize In this paper we discuss some topics on the geometry of type II superstring backgrounds with D-branes, in particular on the geometrical meaning of the D-brane charge, the Ramond-Ramond fields and the Wess-Zumino action. We see that, depending on the behaviour of the D-brane on the four non-compact space-time directions, we need different notions of homology and cohomology to discuss the associated fields and charge: we give a mathematical definition of such notions and show their physical applications. We then discuss the problem of corretly defining Wess-Zumino action using the theory of $p$-gerbes. Finally, we recall the so-called $*$-problem and make some brief remarks about it.

\vskip2cm

\vskip1.5\baselineskip

\vfill
 \hrule width 5.cm
\vskip 2.mm
{\small 
\noindent }
\begin{flushleft}
ferrari@sissa.it
\end{flushleft}
\end{titlepage}

\large

\newtheorem{Theorem}{Theorem}[section]
\newtheorem{Lemma}[Theorem]{Lemma}
\newtheorem{Corollary}[Theorem]{Corollary}
\newtheorem{Rmk}[Theorem]{Remark}
\newtheorem{Def}{Definition}[section]
\newtheorem{ThmDef}[Theorem]{Theorem - Defintion}

\tableofcontents

\newpage

\section{Introduction}

Although D-branes and Ramond-Ramond fields are very familiar objects in string theory, their exact geometrical nature is quite complicated and not so easy to describe. The most natural approach to study superstring backgrounds with D-branes is to use the language of homology and cohomology in the same way it is used for electromagnetism, so that the theory of D-branes becomes actually a generalized version of electromagnetism with higher dimensional sources. In particular, one considers D-branes as sources for violation of Bianchi identity for the Ramond-Ramond fields, so that a Dp-brane charge is the analogue of the magnetic charge for the associated Ramond-Ramond field $G_{8-p}$ and of the electric charge for its Hodge-dual $G_{p+2}$, assuming the democratic formulation of supergravity. Moreover, the Wess-Zumino action, i.e. the minimal coupling of a Dp-brane to the Ramond-Ramond potential $C_{p+1}$, is the analogue of the Wilson line for a charged particle moving in a background electromagnetic field. In order to describe D-brane charges, we can consider D-branes which cover part or all of the non-compact space directions. In this case, the brane cannot be seen as an ordinary homology cycle since, by definition of homology, all cycles have compact support. Thus, in order to correctly describe a theory of electromagnetism with non-compact sources, we are forced to consider a different version of homology, called Borel-Moore homology, which takes into account also non-compact cycles. We will also introduce some modified versions of Borel-Moore homology, in order to describe the possible kinds of D-branes.

Moreover, we deal with the problem of giving a correct definition of Wess-Zumino action. The potential $C_{p+1}$ is a connection on a $p$-gerbe, so that we briefly recall the theory of gerbes with connection, using the language of $\rm\check{C}$ech hypercohomology, in order to explain the meaning of the integral and, if necessary, of its conditions at infinity.

As is well known, this picture is affected by some problems: in particular, magnetic charge is quantized by Dirac quantization condition, and this implies that both a Ramond-Ramond field and its Hodge-dual are quantized, while in general Hodge duality does not preserve quantization. In particular, in type IIB theory the field $G_{5}$ is self-dual, being a D3-brane both electric and magnetic charge with respect to it: but there is not a good lagrangian description of a theory with a source which is at the same time magnetic and electric, and the geometrical translation of this is that it is not possible to quantize a self-dual form and define a correct Wess-Zumino action. These problems have been widely treated in the literature (see \cite{MW} and references therein, \cite{BM2}). In this paper we briefly show a possible non Lorentz-invariant solution, suggested in \cite{Evslin}, which consists in assuming that time and space are orthogonal with respect to the fixed background metric so that one can divide magnetic and electric part of Ramond-Ramond fields and quantize only the magnetic one.

As explained in \cite{MM}, \cite{Witten}, \cite{MW} and \cite{Evslin}, K-theory is a better tool to describe D-brane charges and Ramond-Ramond fields than homology, although it still has some problems. Moreover, the Wess-Zumino action must be completed adding gauge and gravitational couplings (see \cite{MM}). When we consider also such couplings the gerbe description is actually incomplete, and the problem of Dirac quantization becomes more subtle. We will consider the geometrical discussion of this coupling in a future work. Moreover, in all the paper we assume that the $H$-flux is vanishing. We will recall in the conclusions the problems arising with a non-zero $H$-flux.

The paper is organized as follows. In section 2 we recall the basic facts about D-brane charges and Ramond-Ramond fields in type II superstring backgrounds, showing explicitely the analogy with classical electromagnetism. In section 3 we describe Borel-Moore homology and cohomology and we introduce the modifications we will need in the following. In section 4 we use Borel-Moore homology to describe D-branes with any behaviour on the non-compact space-time directions, and also to describe charges directly from the world-volume, as we will discuss in detail. In section 5 we discuss the geometrical meaning of Wess-Zumino action using the theory of $p$-gerbes. In section 6 we recall the so-called $*$-problem and show a possible non Lorentz-invariant solution.

\section{D-brane charge}

We now want to discuss the D-brane charge from the homological point of view. Since this is a generalizations of electromagnetism theory with higher-dimensional sources, we start with a brief review of classical electromagnetism theory in four dimensions. For details the reader can see \cite{Naber}.

\subsection{Preliminaries of electromagnetism}

Let us consider an empty Minkowskian space-time $\mathbb{R}^{1,3}$. Then Maxwell equations are:
\begin{equation}\label{MaxwellEmpty}
	dF = 0 \qquad d*F = 0
\end{equation}
whose solutions represent electric and magnetic fields without sources. In particular, in a fixed reference frame:
\begin{equation}\label{FMatrix}
	F = \begin{bmatrix} 0 & E^{1} & E^{2} & E^{3} \\ -E^{1} & 0 & B^{3} & -B^{2} \\ -E^{2} & -B^{3} & 0 & B^{1} \\ -E^{3} & B^{2} & -B^{1} & 0
	\end{bmatrix}
\end{equation}
and equations \eqref{MaxwellEmpty} assume their classical form $\nabla \times \underline{E} + \frac{\partial \underline{B}}{\partial t} = 0$ and $\nabla \cdot \underline{B} = 0$ for $dF = 0$, and $\nabla \times \underline{B} - \frac{\partial \underline{E}}{\partial t} = 0$ and $\nabla \cdot \underline{E} = 0$ for $d*F = 0$. Since $\mathbb{R}^{1,3}$ is contractible so that the cohomology is zero, both $F$ and $*F$ are exact: $F = dA$, where $A$ is the scalar potential, i.e. $A = (V, \underline{A})$ with $\underline{E} = -\frac{\partial \underline{A}}{\partial t} - \nabla V$ and $\underline{B} = \nabla \times \underline{A}$. Similarly we can find a potential $A'$ such that $*F = dA'$, satisfying the same equations replacing $\underline{B}$ by $\underline{E}$ and $\underline{E}$ by $-\underline{B}$: electric and magnetic fields are interchangable by Hodge-duality, in fact the matrix representation of $*F$ can be obtained from \eqref{FMatrix} again replacing $\underline{B}$ by $\underline{E}$ and $\underline{E}$ by $-\underline{B}$ (the minus is due to the fact that $**F = -F$ in the Minkowskian signature). Thus, up to exchange $F$ and $*F$, electric and magnetic fields without sources are are equivalent.

We now consider an electric charge $q$, moving without accelerating, as a source for the electric field. In this case, Maxwell equations becomes:
\begin{equation}\label{MaxwellCharge}
	dF = 0 \qquad d*F = q \cdot \delta(w)
\end{equation}
where $w$ is the world-line of the particle. This means that we interpret $*F$ not as a form any more but as a current, which is singular in $w$, while in $\mathbb{R}^{1,3} \setminus w$ it is regular and, by equations \eqref{MaxwellCharge}, closed. Instead, $F$ is a closed current on all $\mathbb{R}^{1,3}$, thus it is also exact. Since $H^{2}_{dR}(\mathbb{R}^{1,3} \setminus w) \simeq \mathbb{R}$ ($\mathbb{R}^{1,3} \setminus w$ being homotopic to $S^{2}$), the form $*F$ is in general not exact, actually, as it follows from equations \eqref{MaxwellCharge}, if we consider a linking surface $S^{2} \subset \mathbb{R}^{1,3}$ of $w$ we have that $\int_{S^{2}} *F = q$, thus $[*F]_{dR} \simeq q$ under the isomorphism $H^{2}_{dR}(\mathbb{R}^{1,3} \setminus w) \simeq \mathbb{R}$. In contrast $F$, being exact on the whole $\mathbb{R}^{1,3}$, is exact also when restricted to $\mathbb{R}^{1,3} \setminus w$, so that it is topologically trivial. That's the well-known fact that the electric charge, represented by $F$, is not topological, while the magnetic charge, which is the electric one for $*F$, is encoded in the topology of space-time. Here we see the difference between electric and magnetic charges. In particular, considering a charged particle moving in this background, its actions minimally couples to a potential $A$ of $F$ if we consider the field as electric, in which case $A$ can be globally defined, or to a potential $A'$ of $*F$ (in $\mathbb{R}^{1,3} \setminus w$) if we consider the field as magnetic, in which case $A$ is only local and we must consider gauge transformations (or viceversa if we exchange $F$ and $*F$ up to a sign). In particular the Dirac quantization condition, i.e. the condition that $q \in \mathbb{Z}$ up to a normalization constant, is imposed only on magnetic fields, not on electric ones.

The solutions of \eqref{MaxwellCharge} can all be obtained from a particular one adding the solutions of \eqref{MaxwellEmpty}. One particular solution of \eqref{MaxwellCharge}, in a reference frame in which the charge is fixed in the origin so that $w = \mathbb{R} \times \{0\}$, is:
	\[F = dA, \quad A = \textstyle-\frac{q}{r}dt
\]
where $r$ is the distance of a point from the origin in $\mathbb{R}^{3}$ (thus $A$ is constant in time). Note that the potential $A$ is a $L^{1}_{loc}$-form an all $\mathbb{R}^{3}$, thus $F$ is an exact current in $\mathbb{R}^{3}$. In this way, calling $\vol_{S^{2}} := x_{1}dx_{2} \wedge dx_{3} - x_{2}dx_{1} \wedge dx_{3} + x_{3}dx_{1} \wedge dx_{2}$ the 2-form on $\mathbb{R}^{3}$ restricting to the volume form on $S^{2} \subset \mathbb{R}^{3}$, we get:
	\[\begin{array}{ll}
	& F = dA = \frac{q}{r^{2}} dr \wedge dt = \frac{q}{r^{3}} rdr \wedge dt\\
	& *F = \frac{q}{r^{3}}\vol_{S^{2}}\\
	\textnormal{For $r \neq 0$: } & d*F = -\frac{3q}{r^{4}}dr\wedge \vol_{S^{2}} + \frac{q}{r^{3}}3\vol_{\mathbb{R}^{3}} = \frac{3q}{r^{4}}(r\vol_{\mathbb{R}^{3}} - dr \wedge \vol_{S^{2}}) = 0 \; .
\end{array}\]
Instead, as a current in the whole $\mathbb{R}^{3}$, $d*F = q \cdot \delta(0)$, since:
	\[\begin{split}
	 \langle d*F, \varphi \rangle &= -\langle *F, d\varphi \rangle = -q \int_{\mathbb{R}^{3}} \frac{1}{r^{3}} \vol_{S^{2}} \wedge d\varphi \\
	 & = -q \int_{\mathbb{R}^{3}} \frac{1}{r^{3}} r \frac{d\varphi}{dr}\vol_{\mathbb{R}^{3}} = -q\int_{0}^{+\infty} \frac{d\varphi}{dr}dr = q \cdot \varphi(0)
\end{split}\]
up to a normalization constant. This soluton is static.

\paragraph{}We now make some topological remarks. We consider the following cohomology groups for a manifold $X$:
\begin{itemize}
	\item $H^{n}_{dR}(X)$ is the $n$-th de Rham cohomology group, i.e. the group of closed $n$-forms up to the exact ones;
	\item $H^{n}_{crn}(X)$ is the $n$-th de Rham cohomology group of \emph{currents} on $X$; 
	\item $H^{n}(X, \mathbb{R})$ is $n$-th cohomology group of singular cochains with real coefficients.
\end{itemize}
These three groups are canonically isomorphic. In particular, the natural map $H^{n}_{dR}(X) \rightarrow H^{n}_{crn}(X)$, obtained by thinking of a form as a current, is a canonical isomorphism. To realize an isomorphism between $H^{n}(X, \mathbb{R})$ and $H^{n}_{dR}(X)$ we can use iteratively the Poincar\'e lemma, as explained in \cite{BFS}. For all of these three groups we can consider the compactly-supported version, which we call respectively $H^{n}_{dR,cpt}(X)$, $H^{n}_{crn,cpt}(X)$ and $H^{n}_{cpt}(X, \mathbb{R})$. They are still isomorphic via the restrictions of the previous isomorphisms.

We can define the singular cohomology groups with integral coefficients $H^{n}(X, \mathbb{Z})$, and there is a natural map (not injective in general) $H^{n}(X, \mathbb{Z}) \rightarrow H^{n}(X, \mathbb{R})$ whose image consists of real cohomology classes satisfying Dirac's charge quantization condition: the latter correspond in the de-Rahm cohomology to the forms which give an integral value when integrated over a cycle. Poincar\'e duality provides \emph{on a manifold} a canonical isomorphism $\PD: H_{n}(X, \mathbb{Z}) \overset{\simeq}\longrightarrow H^{\dim(X) - n}_{cpt}(X, \mathbb{Z})$ with the analogous version for real coefficients.

Coming back to the electric source in $\mathbb{R}^{1,3}$, if we restrict the second equation of \eqref{MaxwellCharge} to a fixed instant of time, we get $[d(*F)\vert_{\{t\} \times \mathbb{R}^{3}}]_{cpt} = q \cdot \delta(\{p\})$, for $p = w \cap (\{t\} \times \mathbb{R}^{3})$. Since the point $p$ is compact (contrary to $w$), it defines an homology class $[p] \in H_{0}(\mathbb{R}^{3}, \mathbb{Z})$, thus we can define a compactly-supported cohomology class $\PD_{\mathbb{R}^{3}}([p])$. Under the isomorphism $H^{n}_{cpt}(X, \mathbb{R}) \simeq H^{n}_{crn,cpt}(X)$ one has $\PD_{\mathbb{R}^{3}}([p]) \simeq [\delta(p)]$, hence we obtain from Maxwell equations:
\begin{equation}\label{MaxwellCohomology}
	[d(*F)\vert_{\{t\} \times \mathbb{R}^{3}}]_{cpt} = q \cdot \PD_{\{t\} \times \mathbb{R}^{3}}([p]) \; .
\end{equation}
This identity seems meaningless because we are identifying the class of an exact form with a cohomology class which is in general non-trivial. Actually, we are dealing with \emph{compactly supported} cohomology classes, which can be trivial when considered as generic cohomology classes. Thus, the identity is meaningful and implies that the support of $(*F)\vert_{\{t\} \times \mathbb{R}^{3}}$ is not compact. In this way, we can see the electric (or magnetic) source as a homology cycle conserved in time whose coefficient is the charge; its Poincar\'e dual measures the non-closureness of the associated magnetic field strength as a current.\footnote{We can consider an inertial reference frame in which the charge is fixed in the origin, so that we consider the electric field it creates in $\mathbb{R}^{1,3} \setminus \{(t, 0, 0, 0)\}$. In ths such a frame we can choose the solution of \eqref{MaxwellCharge} given by $\underline{E} = \frac{q}{r^{2}}\underline{u}_{0}$ and $\underline{B} = 0$. As it follows from \eqref{FMatrix}, $F$ is then of the form $F = dt \wedge F'$, thus $*F$ is time-independent and its restriction to any space-slice is the same. In this way we can simply write $d*F = q\cdot \delta(0)$ and $[d*F]_{cpt} = \PD_{\mathbb{R}^{3}}[\{0\}]$, but this picture is not Lorentz invariant.} This viewpoint seems redundant for a point-charge, but for an extended object as a D-brane, which can be topologically non-trivial, it is much more natural.

The cohomological expression \eqref{MaxwellCohomology} is not Lorentz-invariant, since we must fix an instant of time. If we were able to treat $w$ as a homology cycle, we could get from Maxwell equations a Lorentz-invariant expression:
	\[[d*F] = q \cdot \PD_{\mathbb{R}^{1,3}}(w)
\]
without fixing a particular reference frame. We will develop the suitable homology theory to do this.

\paragraph{}We remark that since the de-Rahm cohomology and the cohomology of currents are isomorphic, we can also think of $(*F)_{0}$ as a compactly-supported form whose support is contained in a small neighborhood of the origin. Similarly, the whole $*F$ is a form whose support is contained in a small neighborhood of $t_{0}$. In this case, when we compute the charge as $q = \int_{S^{2}} *F$, we must take $S^{2}$ outside the neighborhood. Using currents or forms is not important, since their cohomology are canonically isomorphic; what really counts is that we consider compactly-supported classes, which can be non-trivial also in $\mathbb{R}^{3}$. However, it is more natural to use currents since Maxwell equations are naturally formulated with a $\delta$-function.

\subsection{Charge of a D-brane}

We consider type II superstring theory in a ten-dimensional space-time of the form $S = \mathbb{R}^{1,3} \times X$ for $X$ in general compact but not necessarily, such that the background metric in $\mathbb{R}^{1,3}$ is the standard Minkowskian metric $\eta^{\mu\nu}$ and \emph{the $H$-flux is zero}. A Dp-brane $Y_{p}$ has a $(p+1)$-dimensional world-volume $WY_{p} \subset S$, which represents a classical trajectory in space-time. To define the charge of the D-brane, as for a particle we think that it is moving without accelerating in the non-compact directions $\mathbb{R}^{1,3}$ (so the projection on $X$ is fixed), so that the violated Bianchi identity becomes:
\begin{equation}\label{MaxwellDBrane}
	dG_{8-p} = q \cdot \delta(WY_{p}) \qquad dG_{p+2} = 0
\end{equation}
where $q$ is the charge, or equivalently, the number of D-branes in the stack. To compute the charge from the background data, we consider a linking manifold\footnote{A linking manifold is the boundary of a manifold intersecting $WY_{p}$ tranversally in isolated points of its interior.} $L$ of $WY_{p}$ in $S$ with linking number $l$, so that we have:
	\[q = \frac{1}{l} \int_{L} G_{8-p} \; .
\]
We can always choose a linking sphere (so that $l = 1$) near non-singular points of $WY_{p}$: in fact, we choose near a non-singular point $p \in WY_{p}$ a reference frame such that $WY_{p}$ corresponds to the first $p+1$ coordinates, then we take a small sphere in the transverse cohordinates. From Dirac quantization condition (see section \ref{WessZumino}) we know that the charge is quantized, thus $G_{8-p}$ must be an integral form. In particular, since by \eqref{MaxwellDBrane} we see that $G_{8-p}$ is not closed, we should say that $G_{8-p}$ restricted to the complement of $WY_{p}$ represents an integral cohomology class.

We now suppose that the brane is a particle in $\mathbb{R}^{1,3}$. In a fixed reference frame we call $M$ the space manifold $M = \mathbb{R}^{3} \times X$. We fix at an instant of time $t$ the D-brane volume $Y_{p,t} \subset \{t\} \times M$. We call $M_{t} := \{t\} \times M$. Then, the violated Bianchi identity becomes $d_{M_{t}}(G_{8-p}\vert_{M_{t}}) = q \cdot \delta(Y_{p,t})$ so that, \emph{if $Y_{p,t}$ is compact} (which is always the case when the brane is a particle in $\mathbb{R}^{1,3}$ if $X$ is compact), we obtain:
\begin{equation}\label{BianchiId}
	[\, d_{M_{t}}(G_{8-p}\vert_{M_{t}}) \,]_{\cpt} = \PD_{M_{t}}(q \cdot Y_{p,t}) \; .
\end{equation}
As pointed out before, it is important that the space-time $M$ is non-compact (see \cite{MW} and \cite{FS}), so that the Poincar\'e dual of the brane volume is a \emph{compactly supported} cohomology class, which can be trivial as a generic cohomology class. Thus, the identity \eqref{BianchiId} implies that, for homologically non-trivial branes, the support of $G_{8-p}\vert_{\{t\} \times M}$ is not compact. In particular, $\PD_{M}(q \cdot Y_{p})$ must live in the kernel of the natural map $\iota: H^{9-p}_{\cpt}(M) \rightarrow H^{9-p}(M)$. We could also write the first equation as $dG_{8-p} = \PD_{S}(q \cdot WY_{p})$, but, since $WY_{p}$ is in general non-compact\footnote{If the brane is stable it exists for all the time, from $-\infty$ to $+\infty$, thus the world-volume is non compact.} and it does not define a homology cycle, we postpone this discussion.

We can compute the charge $q$ at any fixed instant: if we consider a linking surface $L_{t}$ of $Y_{p,t}$ in $M_{t}$ with linking number $l$, we have $q = \frac{1}{l} \int_{L_{t}} (G_{8-p}\vert_{M_{t}})$. The charge $q$ is conserved in time, actually all the homology class of the D-brane is conserved. In fact, let us consider two volumes $Y_{p,t_{1}}$ and $Y_{p,t_{2}}$. Then we can consider the piece of the world-volume linking them, which is $(WY_{p})\vert_{[t_{1}, t_{2}] \times M}$. If we consider the canonical identification $M_{t_{1}} \simeq M_{t_{2}} \simeq M$, we can consider both $Y_{t_{1}}$ and $Y_{t_{2}}$ as cycles in $M$. If we consider the projection $\pi: [t_{1}, t_{2}] \times M \rightarrow M$, then $\pi((WY_{p})\vert_{[t_{1}, t_{2}] \times M})$ is a singular chain in $M$ which makes $Y_{t_{1}}$ and $Y_{t_{2}}$ homologous. Thus they have the same Poincar\'e dual and they define the same charge.

\paragraph{}As for classical electromagnetism, the solutions of \eqref{MaxwellDBrane} can be obtained from a fixed one adding the solution to the equations in the empty space:
\begin{equation}\label{MaxwellDEmpty}
	dG_{8-p} = 0 \qquad dG_{p+2} = 0 \; .
\end{equation}
We study a particular static solution, which we aspect to be similar to the one of classical electromagnetism. Let us consider a brane that is a particle in $\mathbb{R}^{1,3}$ and a reference frame in which it is fixed in the origin. Thus we have a cycle $Y_{p} \subset \{0\} \times X$. We consider the case in which there is a foliation of $(\mathbb{R}^{3} \times X) \setminus Y_{p}$ made by manifolds of points at a fixed distance from $Y_{p}$, as in classical electromagnetism where the origin foliates $\mathbb{R}^{3} \setminus \{0\}$ in spheres: for example, if we imagine a torus embedded in $\mathbb{R}^{3}$ in the standard way and we consider a vertical circle as a cycle, it foliates the torus in couples of circles parallel to it at a fixed distance (with the exception of the opposite one, in which case the two circles of the couple collapse to the same one). We now consider for a point $x \in (\mathbb{R}^{3} \times X) \setminus Y_{p}$ the manifold $Z_{x}$ of the foliation containing $x$, which has dimension 8 (independently on $p$), since to cover a neighborhood of it we need the coordinates on $Z_{x}$ and only one parameter more, the distance from $Y_{p}$. Then, in $T_{x}(Z_{x})$, we consider the subspace $V_{x}$ parallel to the D-brane, i.e., if $d(x, Y_{p}) = d(x,y)$, we consider the submanifold of points in $Z_{x}$ dinstant $r$ from $y$ and we consider its orthogonal. We call $z_{1}, \ldots, z_{p}$ an orthonormal system of generators of $V_{x}$. Then we define:
	\[G_{p+2} = dA, \quad A = -\frac{q}{9-p-2} \cdot \frac{1}{r^{9-p-2}} \, dz_{1} \wedge \ldots \wedge dz_{p} \wedge dt
\]
so that:
	\[\begin{array}{rcl}
	G_{p+2} & = & dA \,=\, \frac{q}{r^{9-p-1}} \, dr \wedge dz_{1} \wedge \ldots \wedge dz_{p} \wedge dt\\
	& = & \frac{q}{r^{9-p}} \, r dr \wedge dz_{1} \wedge \ldots \wedge dz_{p} \wedge dt \\
	G_{8-p} & = & \frac{q}{r^{9-p}} \vol_{Z}
\end{array}\]
where $Z$ is a submanifold of points at fixed distance with respect to $Y_{p}$. In this way, as before, $dG_{8-p} = \delta(Y_{p})$. In this solution $G_{p+2}$ is exact, while $G_{8-p}$ is non-trivial only on the cycle given by a linking sphere of $Y_{p}$ with weigth $q$: all such linking spheres are homologous, since, if they have both the same radius, they are linked by a piece of the suitable leaf $Z$ of the foliation. In the next paragraph we give a more complete discussion of this.

\paragraph{}For what concerns the solutions of \eqref{MaxwellDEmpty}, from a matrix representation analoguous to \eqref{FMatrix} we get usual Maxwell equations. In particular, we split the Ramond-Ramond fields in the following way:
\begin{equation}\label{RRSplit}
	G_{p} = G^{s}_{p} + dt \wedge G^{t}_{p-1}
\end{equation}
so that, calling $*_{9}$ the Hodge-$*$ in $M$, which is Euclidean:
\begin{equation}\label{RRHodge}
	*G_{p} = -*_{9}G^{t}_{p-1} + dt \wedge (-1)^{p}*_{9}G^{s}_{p} \; .
\end{equation}
In fact, all the terms of $G^{s}_{p}$ are of the form $f \cdot dx_{i_{1}} \wedge \ldots \wedge dx_{i_{p}}$, and their Hodge-$*$ is $\varepsilon^{i_{1}, \ldots, i_{p}, 0, j_{1}, \ldots, j_{9-p}} f \cdot dt \wedge dx_{j_{1}} \wedge \ldots \wedge dx_{j_{9-p}} = (-1)^{p}\varepsilon^{i_{1}, \ldots, i_{p}, j_{1}, \ldots, j_{9-p}} f \cdot dt \wedge dx_{j_{1}} \wedge \ldots \wedge dx_{j_{9-p}} = (-1)^{p}dt \wedge *_{9} f \cdot dx_{i_{1}} \wedge \ldots \wedge dx_{i_{p}}$. Similarly, all the terms of $dt \wedge G^{t}_{p-1}$ are of the form $f \cdot dt \wedge dx_{i_{1}} \wedge \ldots \wedge dx_{i_{p-1}}$ and their Hodge-$*$ is $-\varepsilon^{0,i_{1}, \ldots, i_{p-1},j_{1}, \ldots, j_{10-p}} f \cdot dt \wedge dx_{i_{1}} \wedge \ldots \wedge dx_{i_{p-1}} = -\varepsilon^{i_{1}, \ldots, i_{p-1},j_{1}, \ldots, j_{10-p}} f \cdot dt \wedge dx_{i_{1}} \wedge \ldots \wedge dx_{i_{p-1}} = -*_{9}f \cdot dt \wedge dx_{i_{1}} \wedge \ldots \wedge dx_{i_{p-1}}$, the minus sign being due to the fact that $dt$ is negative definite (for a review of Hodge-$*$ with Minkowskian signature see appendix \ref{HodgeMinkowski}). Then the solutions of $dG_{p} = 0$ and $d*G_{p} = 0$ becomes:
	\[\begin{array}{lllll}
	dG_{p} = 0: & & \frac{\partial G^{s}_{p}}{\partial t} - d_{9}G^{t}_{p-1} = 0 & & d_{9}G^{s}_{p} = 0\\
	d*G_{p} = 0: & & -*_{9}\frac{\partial G^{t}_{p-1}}{\partial t} - (-1)^{p} d_{9}*_{9}G^{s}_{p} = 0 & & d_{9}*_{9}G^{t}_{p-1} = 0
\end{array}\]
which correspond to Maxwell equations for $p = 2$ in dimension 3 if we identify $G^{s}_{p} = *_{3}\varphi(B)$ and $G^{t}_{p-1} = -\varphi(E)$ for $\varphi: T(\mathbb{R}^{3}) \rightarrow T^{*}(\mathbb{R}^{3})$ the isomorphism given by the metric.

This is different from classical electromagnetism theory since in contrast with $\mathbb{R}^{1,3}$ which is contractible, the space-time $\mathbb{R}^{1,3} \times X$ can have non-trivial cycles in itself, even before putting the charge sources. Thus, the equations $dG_{8-p} = 0$ and $dG_{p+2} = 0$ do not imply that $G_{8-p}$ and $G_{p+2}$ are exact. We briefly analyze this difference. Since $\mathbb{R}^{1,3}$ is contractible, the natural immersion $i: X \rightarrow \mathbb{R}^{1,3} \times X$, defined by $i(x) = (0, x)$, induces an isomorphism in cohomology $i^{*}: H^{*}_{dR}(\mathbb{R}^{1,3} \times X) \overset{\simeq}\longrightarrow H^{*}_{dR}(X)$ sending a class $[\omega]$ in the class $[\omega_{0}]$ for $\omega_{0} := \omega\vert_{\{0\} \times X}$. Thus, for any closed $p$-form $\omega$, we have $\omega = \omega_{0} + d\rho$ with\footnote{We should write $\omega = \pi^{*}\omega_{0} + d\rho$ for $\pi: \mathbb{R}^{1,3} \times X \rightarrow X$ the projection, but for simplicity we idenfity a form on $X$ (as $\omega_{0}$) with the corresponding form on $\mathbb{R}^{1,3} \times X$ which does not depend on $\mathbb{R}^{3}$ (as $\pi^{*}\omega_{0}$).} $\omega_{0} \in \Lambda^{p}T^{*}X$ \emph{closed} and $\rho \in \Lambda^{p-1}T^{*}(\mathbb{R}^{1,3} \times X)$. Now, since $X$ is compact, we can apply Hodge decomposition theorem (see \cite{GH}) to $\omega_{0}$ so that, being it closed, we obtain $\omega_{0} = h_{0} + d\rho_{0}$ with $h_{0}$ \emph{harmonic} in $X$. We can suppose $d\rho_{0}$ already included in $d\rho$, so we finally get:
\begin{equation}\label{Hodge}
	\omega = h_{0} + d\rho
\end{equation}
with $h_{0} \in \Lambda^{p}T^{*}X$ \emph{harmonic} and $\rho \in \Lambda^{p-1}T^{*}(\mathbb{R}^{1,3} \times X)$. The form $h_{0}$ is uniquely determined by the cohomology class of $\omega$, thus, if we fix such a class, we remain with the freedom of $\rho$. In particular, we have that:
	\[G_{p} = (h_{0})_{p} + d\rho_{p} \qquad *G_{p} = *(h_{0})_{p} + *d\rho_{p}
\]
where $\rho_{p}$ is the analogue of the potential $A$. We remark that $*(h_{0})_{p}$ is exact, as $*h_{0} = (-1)^{p}dt \wedge dx_{1} \wedge dx_{2} \wedge dx_{3} \wedge *_{6}h_{0} = d\bigl( (-1)^{p} x_{1} \wedge dx_{2} \wedge dx_{3} \wedge *_{6}h_{0}\bigr)$, where in the last equality we used the fact that $*_{6}h_{0}$ is closed since $h_{0}$ is harmonic in $X$. Thus, the non-triviality of the space-time topology is encoded in $(h_{0})_{p}$ for the electric charge and in the possible non-triviality of $*d\rho_{p}$ for the magnetic one. We interpret this physically by noting that a non-trivial cycle can be thought of as a trivial one by removing a charge (so that the charge is encoded in the background). For example, in classical electromagnetism if we consider the background $\mathbb{R}^{1,3} \setminus w$ for $w$ the world-line of a charge, in that background Maxwell equations in empty space are satisfied, but the topology is non-trivial and $*F$ is not exact. The present situation is analogous.

\paragraph{}Up to now we have assumed the existence of a suitable foliation of space-time in order to reproduce a situation analogous to the one of classical electromagnetism. In fact we can show that we can solve the Maxwell equations in general. We search a static solution $G_{p} = dt \wedge G^{t}_{p-1}$, so that, thanks to \eqref{RRHodge}, we have $*G_{p} = -*_{9}G^{t}_{p-1}$. We use smooth forms instead of currents for simplicity, then it will be immediate to reduce to Maxwell equations formulated with $\delta$-functions. Let us consider a form $G^{t}_{p-1} \in \Lambda^{p-1}(\mathbb{R}^{1,3} \times X)$, decomposed as in \eqref{Hodge}: we now want to study the compactly-supported cohomology class of $d*_{9}G^{t}_{p-1}$. Given a function $e: \mathbb{R} \rightarrow \mathbb{R}$ such that $\int_{-\infty}^{+\infty} e = 1$, for any manifold $A$ there is an isomorphism:
\begin{equation}\label{IsoCptCohom}
\begin{split}
	e_{*}: \;&H_{dR,\cpt}^{n-1}(A) \overset{\simeq}\longrightarrow H_{dR,\cpt}^{n}(\mathbb{R} \times A)\\
	&[\,\eta\,] \longrightarrow [\,e(x)\,dx \wedge \eta\,]
\end{split}
\end{equation}
whose inverse is the pull-back $\pi^{*}$ of the projection $\pi: \mathbb{R} \times A \rightarrow A$ (see \cite{BT}).\footnote{For currents the isomorphism \eqref{IsoCptCohom} can be described by $[\delta(Y)] \rightarrow [\delta(\{0\} \times Y)]$.} Thus, fixing three functions $e_{1}, e_{2}, e_{3}$ with integral $1$ we obtain an isomorphism $H^{n-3}_{dR}(X) \overset{\simeq}\longrightarrow H_{dR,\cpt}^{n}(\mathbb{R}^{3} \times X)$ given by $[\,\eta\,] \longrightarrow [\,e_{1}(x)\,dx_{1} \wedge e_{2}(x)\,dx_{2} \wedge e_{3}(x)\,dx_{3} \wedge \eta\,]$. If we want to fix the cohomology class $[\,d*G^{t}_{p-1}\,]_{\cpt}$, we can choose $\alpha$ harmonic on $X$ corresponding (uniquely) to the fixed class under the latter isomorphism, and require:
	\[d*G^{t}_{p-1} = e_{1}(x)\,dx_{1} \wedge e_{2}(x)\,dx_{2} \wedge e_{3}(x)\,dx_{3} \wedge \alpha + d\xi_{cpt}
\]
for any compactly-supported form $\xi_{cpt}$. In order to show how to solve this equation, we remark that:
\begin{itemize}
	\item considering \eqref{Hodge}, we have that $d*_{9}h_{0} = 0$, since $*_{9}h_{0} = (-1)^{p}dx_{1} \wedge dx_{2} \wedge dx_{3} \wedge *_{6}h_{0}$ where $*_{6}$ is the Hodge-dual on $X$; hence, being $h_{0}$ harmonic in $X$, $d*_{9}h_{0} = 0$, so that we have to consider the cohomology class $[\,d*_{9}d\rho\,]$;
	\item for $\alpha$ closed, $e(x)\,dx \wedge \alpha = d\bigl(\int_{0}^{x}e \cdot \alpha\bigr)$ as one can see from the Leibnitz rule or directly from the definition of exterior differential.
\end{itemize}
Thus we obtain:
	\[\begin{array}{l}
	d*_{9}d\rho = e_{1}(x)\,dx_{1} \wedge e_{2}(x)\,dx_{2} \wedge e_{3}(x)\,dx_{3} \wedge \alpha + d\xi_{cpt}\\
	d*_{9}d\rho = d\bigl(\int_{0}^{x_{1}}e_{1} \cdot e_{2}(x)\,dx_{2} \wedge e_{3}(x)\,dx_{3} \wedge \alpha\bigr) + d\xi_{cpt}\\
	*_{9}d\rho = \int_{0}^{x_{1}}e_{1} \cdot e_{2}(x)\,dx_{2} \wedge e_{3}(x)\,dx_{3} \wedge \alpha + \xi_{cpt} + \eta_{closed}\\
	d\rho = \int_{0}^{x_{1}}e_{1} \cdot *_{9}\bigl(e_{2}(x)\,dx_{2} \wedge e_{3}(x)\,dx_{3} \wedge \alpha\bigr) + *_{9}\xi_{cpt} + *_{9}\eta_{closed} \; .
\end{array}\]
Let us show that the first term of the r.h.s. is actually exact. Since $*_{9}\bigl(e_{2}(x)\,dx_{2} \wedge e_{3}(x)\,dx_{3} \wedge \alpha\bigr) = (-1)^{p-1}dx_{1} \wedge *_{6}\alpha$ we obtain:
	\[\begin{split}
	\textstyle \int_{0}^{x_{1}}e_{1} \cdot *_{9}\bigl(e_{2}(x)\,dx_{2} \wedge e_{3}(x)\,dx_{3} \wedge \alpha\bigr) &= \textstyle (-1)^{p} (\int_{0}^{x_{1}}e_{1})dx_{1} \wedge *_{6}\alpha\\
	&= \textstyle d\bigl( (-1)^{p} \int_{0}^{x_{1}} \int_{0}^{y_{1}} e_{1} \cdot *_{6}\alpha \bigr)
\end{split}\]
where the last equality is due to the fact that $*_{6}\alpha$ is closed since $\alpha$ has been chosen harmonic on $X$. Hence we obtain:
\begin{equation}\label{Rho}
	\textstyle \rho = (-1)^{p} \int_{0}^{x_{1}} \int_{0}^{y_{1}} e_{1} \cdot *_{6}\alpha + \psi + \lambda_{closed}
\end{equation}
where $d\psi = *_{9}\xi_{cpt} + *_{9}\eta_{closed}$. The form $\psi$, in particular for what concers $\eta_{closed}$, encodes the freedom of Maxwell equations in empty space.

Asking $[d*_{9}G^{t}_{p-1}]_{cpt} = [e_{1}(x)\,dx_{1} \wedge e_{2}(x)\,dx_{2} \wedge e_{3}(x)\,dx_{3} \wedge \alpha]_{cpt}$, we found no obstructions on $\alpha$: this could seem strange, since the r.h.s. must represent a class which is exact in the ordinary cohomology (not compactly supported), being $d*_{9}\omega$ exact. In particular, $[d*_{9}\omega]_{\cpt}$ lies in the kernel of the natural map $\iota: H^{9-p}_{\cpt}(\mathbb{R}^{3} \times X) \rightarrow H^{9-p}(\mathbb{R}^{3} \times X)$. Actually there is no contradiction, since, for manifolds of the form $\mathbb{R} \times A$, the map $\iota$ is the zero map, i.e., every \emph{closed} compactly-supported form on $\mathbb{R} \times A$ is exact, although not necessarily compactly-supported exact. In fact, considering the isomorphism \eqref{IsoCptCohom}, we see that every class in $H^{p}_{\cpt}(\mathbb{R} \times A)$ is represented by $e(x)\,dx \wedge \eta$ for $\eta$ closed, and, as we have already shown, $e(x)\,dx \wedge \eta = d\bigl(\int_{0}^{x}e \cdot \eta\bigr)$. We can also see that $\iota = 0$ considering the following maps:
	\[H^{*-1}_{\cpt}(A) \overset{e_{*}}\longrightarrow H^{*}_{\cpt}(\mathbb{R} \times A) \overset{\iota}\longrightarrow H^{*}(\mathbb{R} \times A) \overset{i^{*}}\longrightarrow H^{*}(A) \;.
\]
The composition is the zero map, since, for a fixed form $\eta$, being $\iota$ the idendity on the representative, the composition is $i^{*}(e(x)\,dx \wedge \eta) = (e(x)\,dx \wedge \eta)\vert_{\{0\} \times A}$, but the restriction of $e(x)\,dx$ gives $0$. Since $e_{*}$ and $i^{*}$ are isomorphisms, the only possibility is that $\iota = 0$.

This shows that, fixing the class of $Y_{p}$, we can always solve \eqref{BianchiId}. To obtain exactly \eqref{MaxwellDBrane} we use modify $\xi_{\cpt}$ with a current whose differential is the difference between $\delta(Y_{p})$ and the form $d*G^{t}_{p-1}$ obtained with the previous procedure.

\subsection{Summary}

Summarizing, for a Dp-brane with world volume $WY_{p}$ we have equations:
	\[dG_{8-p} = q \cdot \delta(WY_{p}) \qquad dG_{p+2} = 0
\]
from which we obtain the cohomological relation:
	\[[\, d_{M_{t}}(G_{8-p}\vert_{M_{t}}) \,]_{\cpt} = \PD_{M_{t}}(q \cdot Y_{p,t})
\]
and we compute the charge as:
	\[q = \frac{1}{l} \int_{L} G_{8-p}
\]
for $L$ a linking manifold of $WY_{p}$ in $S$. The solutions of this system are given by one particular solution, which under suitable hypoteses is similar to the static one for classical electromagnetism, and a generic solution of the equations in empty space. The particular solution can be obtained by an exact electric field strength and a magnetic one which is non-trivial only on the cycle obtained removing the charge, while the solutions in empty space can add topologically non-trivial terms in any cycle. We interpret these terms as charges hidden in the hole of the cycles which are not considered in our space-time region.

The questions we would like to address in this picture concern the cohomogical equations. In particular:
\begin{itemize}
	\item we must assume that the brane volume is \emph{compact} at any instant of time, thus, e.g. for $S = \mathbb{R}^{1,3} \times X$, the brane must be a particle in the non-compact directions $\mathbb{R}^{1,3}$; in the other cases we cannot describe the D-brane charge as a homology cycle conserved in time;
	\item the equations are not Lorentz-invariant, since the whole world-volume is non-compact and we cannot have a global formulation.
\end{itemize}
The second question arises also in classical electromagnetism, since the world-line of a particle is not compact, while the first is specific to D-brane theory. We now introduce suitable homology and cohomology theories in order to solve these problems.

\section{Borel-Moore homology}

In the ordinary singular homology any cycle must be compact. However, there is a suitable notion of homology, called Borel-Moore homology (see \cite{BM}), which takes into account also non-compact cycles, and, as we now show, it naturally appears if we start from cohomology and we want to define Poincar\'e duals of non compactly-supported classes. It is usually treated in the literature in the sheaf-theoretic or simplicial version, thus we give a description analogous to the one of singular homology. We first briefly recall the definition of singular homology and cohomology (for details see \cite{Hatcher}) to compare it with the Borel-Moore one.

\subsection{Definition of Borel-Moore homology}

We denote by $\Delta^{n} = \{x \in \mathbb{R}^{n+1}: x_{1} + \cdots + x_{n+1} = 1, \, x_{i} \geq 0 \,\forall i\}$ the $n$-dimensional simplex with the Euclidean topology. For $0 \leq k \leq n$, we denote by $(\Delta^{n})^{k}$ the $k$-th face of $\Delta^{n}$ obtained ``removing'' the $k$-vertex, i.e. $(\Delta^{n})^{k} = \Delta^{n} \cap \{x: x_{k+1} = 0\}$. Given a topological space $X$, we consider its set of \emph{$n$-chains} defined as the free abelian group generated by continuous maps from $\Delta^{n}$ to $X$:
	\[C_{n}(X, \mathbb{Z}) := \bigoplus_{\{\sigma^{n}: \Delta^{n} \rightarrow X\}} \mathbb{Z}
\]
and we define a boundary operator $\partial_{n}: C_{n}(X, \mathbb{Z}) \rightarrow C_{n-1}(X, \mathbb{Z})$ given by:
\begin{equation}\label{Boundary}
	\partial_{n}(\sigma^{n}) := \sum_{k=0}^{n} (-1)^{k} \bigl(\sigma^{n} \circ i_{k}^{n-1}\bigr)
\end{equation}
where $i_{k}^{n-1}: \Delta^{n-1} \rightarrow (\Delta^{n})^{k}$ is the standard linear immersion. One can prove that $\partial_{n-1} \circ \partial_{n} = 0$, so that one can define the \emph{singular homology groups} of $X$ as:
	\[H_{n}(X, \mathbb{Z}) := \Ker \,\partial_{n} \,/\, \IIm \,\partial_{n+1} \; .
\]

\paragraph{}Given an $n$-chain $\sum_{\sigma^{n}} k_{\sigma^{n}}$, we define its \emph{support} as the union of the images of the $n$-simplices with non-zero coefficient, i.e. $\bigcup_{\sigma^{n} \,\vert\, k_{\sigma^{n}} \neq 0} \sigma^{n}(\Delta^{n})$. The fact that only finitely many coefficients are non-zero implies that the support of any chain is \emph{compact}. In particular, the support of a cycle is compact without boundary.\footnote{In general the support is not a manifold, it can have singularities. Actually, it can happen that there are homology classes in a smooth manifold which have no representatives made by smooth submanifolds (see \cite{BHK}).} Thus, for example in $\mathbb{R}^{2}$, the circle $S^{1}$ is the support of some homology cycles (for example, the one obtained triangulating $S^{1}$ with two half-circles), but an infinite line, e.g. one of the two coordinate-axes, is not.

There is a different version of homology, called \emph{Borel-Moore homology}, which takes into account also non-compact cycles. To define it, one might think that the right solution is to define chains using direct product instead of direct sum (the difference between direct sum and direct product is briefly recalled in appendix \ref{DirectSumProd}), but in this way we would have no control on the geometry of their support: for example, any subset $A \subset X$, also very irregular, should be the support of a $0$-chain, e.g. the one defined giving the coefficient $1$ to the points of $A$ and $0$ to the points of $X \setminus A$. Moreover, in this case we could not define the boundary operator: in fact, let us suppose in $\mathbb{R}^{2}$ to give coefficeint $1$ to the $1$-simplices made by the rays of the disc $D^{2}$ (or to infinitely many of them, not necessarily all), and $0$ to all the others. In this case, applying the boundary operator, the origin should have infinite coefficient, thus the boundary is not well-defiend. We thus need some conditions. We give the following definitions:
\begin{Def} $ $
\begin{itemize}
	\item A \emph{generalized $n$-chain} on a topological space $X$ is an element of the direct product:
	\[C'_{n}(X, \mathbb{Z}) := \prod_{\{\sigma^{n}: \Delta^{n} \rightarrow X\}} \mathbb{Z} \; .
\]
	\item The \emph{support} of a generalized $n$-chain $\prod_{\sigma^{n}} k_{\sigma^{n}}$ is $\bigcup_{\sigma^{n} \,\vert\, k_{\sigma^{n}} \neq 0} \sigma^{n}(\Delta^{n})$.
	\item A generalized $n$-chain $\prod_{\sigma^{n}} k_{\sigma^{n}}$ is called \emph{locally finite} if for every $x \in X$ there exists a neighborhood $U \subset X$ of $x$ such that there exist only finitely many simplices $\sigma^{n}$ with non-zero coefficient whose image has non-empty intersection with $U$.
\end{itemize}
\end{Def}
On locally finite chains we can correctly define the boundary operator. In fact, let us consider such a chain $\prod_{\sigma^{n}} k_{\sigma^{n}}$ and let us fix an $(n-1)$-simplex $\tilde{\sigma}^{n-1}$ which lie in the boundary of some $\sigma^{n}$ with non-zero coefficient: we show that it must lie in the boundary of only finitely many of them. In fact, for every $p$ in the image of $\tilde{\sigma}^{n-1}$ we choose a neighborhood provided by the local finiteness condition. Since the image is compact, we can select only finitely many such neighborhoods. We have thus found a neighborhood of the image of $\tilde{\sigma}^{n-1}$ which intersects only finitely many simplices $\sigma^{n}$ with non-zero coefficient: since any simplex intersects its boundary, only finitely many $\sigma^{n}$-s can have $\tilde{\sigma}^{n-1}$ as boundary, so that we have no obstructions in extending the boundary operator also to infinite sums of this kind. We can now define Borel-Moore singular homology.
\begin{Def} $ $
\begin{itemize}
	\item A \emph{Borel-Moore $n$-chain} is a generalized $n$-chain which is \emph{locally finite} and has \emph{closed support}.
	\item Calling $\partial_{n}^{BM}$ the boundary operator extended to locally finite generalized $n$-chains and restricted to Borel-Moore ones, we define the \emph{Borel-Moore singular homology groups} as:
		\[H_{n}^{BM}(X, \mathbb{Z}) := \Ker \,\partial_{n}^{BM} \,/\, \IIm \,\partial_{n+1}^{BM} \; .
\]
\end{itemize}
\end{Def}
Let us consider $\mathbb{R}^{2}$ and a Borel-Moore cycle whose support is a line, e.g the $x$-axis with a suitable triangulation. Of course it is not a cycle in ordinary homology, but if we add a point at infinity, i.e. we compactify $\mathbb{R}^{2}$ to $S^{2}$, the line becomes a circle in $S^{2}$, thus a cycle in ordinary homology. This is a general fact, actually one can prove that, for $X^{+}$ the one-point compactification of a space $X$, there is a canonical isomorphism $H_{n}^{BM}(X, \mathbb{Z}) \simeq H_{n}((X^{+}, \{\infty\}), \mathbb{Z})$. Under suitable hypotesis of regularity (i.e. that $\{\infty\}$ is closed and a deformation retract of one of its neighborhoods, which always happens if $X^{+}$ is a manifold), $H_{n}((X^{+}, \{\infty\}), \mathbb{Z}) \simeq \tilde{H}_{n}(X^{+}, \mathbb{Z})$. Thanks to this isomorphism we can compute more easily the Borel-Moore homology groups.

\paragraph{}We now see some examples, comparing Borel-Moore homology with the ordinary one. For $\mathbb{R}^{n}$:
	\[H_{n}^{BM}(\mathbb{R}^{n}, \mathbb{Z}) = \mathbb{Z} \qquad H_{k}^{BM}(\mathbb{R}^{n}, \mathbb{Z}) = 0 \; \forall k \neq n \; .
\]
This immediately follows from that fact that $(\mathbb{R}^{n})^{+} \simeq S^{n}$ so that $H_{k}^{BM}(\mathbb{R}^{n}, \mathbb{Z}) \simeq \tilde{H}_{k}(S^{n}, \mathbb{Z})$. We know that for ordinary homology the only non-zero group is $H_{0}(\mathbb{R}^{n}, \mathbb{Z}) = \mathbb{Z}$. The non-trivial cycle in $H_{n}^{BM}(\mathbb{R}^{n}, \mathbb{Z})$ is the whole $\mathbb{R}^{n}$ itself: if we consider an infinite triangulation of $\mathbb{R}^{n}$ and we give coefficient $1$ to each simplex of the triangulation we describe it as a Borel-Moore cycle, and one can show that it is not a boundary. For ordinary homology it is not a cycle since it is non-compact. Moreover, the origin (or any other point) is a non-trivial cycle in ordinary homology, that's why $H_{0}(\mathbb{R}^{n}, \mathbb{Z}) = \mathbb{Z}$. This cycle becomes trivial in Borel-Moore homology: in fact, a half-line from the origin to infinity is a $1$-chain whose boundary is exactly the origin,\footnote{One may wonder why the origin becomes trivial in the Borel-Moore homology while, even in the one-point compactification, it remains a non-trivial cycle. The point is that to realize the isomorphism $H_{0}^{BM}(\mathbb{R}^{n}, \mathbb{Z}) \simeq \tilde{H}_{0}((\mathbb{R}^{n})^{+}, \mathbb{Z})$ a cycle in the Borel-Moore homology of $\mathbb{R}^{n}$ becomes a cycle in $(\mathbb{R}^{n})^{+}$ adding the point at infinty, as for the $x$-axis that we considered. Thus, to the origin of $\mathbb{R}^{n}$ we must also add the point at infinity: we thus obtain a couple of points in $S^{n}$, which is the boundary of the segment linking them, and such a segment is exactly the completion of the half-line trivializing the origin in $\mathbb{R}^{n}$.} that's why $H_{0}^{BM}(\mathbb{R}^{n}, \mathbb{Z}) = 0$.

As another example we compute Borel-Moore homology of $\mathbb{R}^{n} \setminus \{0\}$. For this we use another isomorphism, since the one-point compactification is not a good space: if $\overline{X}$ is any compactification of $X$, under suitable hypoteses there is a canonical isomorphism $H_{n}^{BM}(X, \mathbb{Z}) \simeq H_{n}((\overline{X}, \overline{X} \setminus X), \mathbb{Z})$. We thus consider $X = \mathbb{R}^{n} \setminus \{0\}$ and $\overline{X} = S^{n}$ and we call $\overline{X} \setminus X = \{N, S\}$ thinking to north and south poles. We thus have to compute $H_{k}((S^{n}, S^{n} \setminus \{N, S\}), \mathbb{Z})$. We consider the long exact sequence:
	\[\xymatrix{
	\cdots \ar[r] & H_{k}(\{N, S\}) \ar[r] & H_{k}(S^{n}) \ar[r] & H_{k}(S^{n}, \{N, S\}) \ar[r] & H_{k-1}(\{N, S\}) \ar[r] & \cdots
}\]
We suppose $n \geq 2$. Then, for $k \geq 2$ the sequence becomes:
	\[\xymatrix{
	\cdots \ar[r] & 0 \ar[r] & H_{k}(S^{n}) \ar[r] & H_{k}(S^{n}, \{N, S\}) \ar[r] & 0 \ar[r] & \cdots
}\]
so that $H_{k}^{BM}(\mathbb{R}^{n} \setminus \{0\}, \mathbb{Z}) \simeq H_{k}(S^{n})$, i.e. $\mathbb{Z}$ for $k = n$ and $0$ for $2 \leq k \leq n-1$. This is different from ordinary homology in which, being $\mathbb{R}^{n} \setminus \{0\}$ homotopic to $S^{n-1}$, we have $H_{k}(\mathbb{R}^{n} \setminus \{0\}, \mathbb{Z}) = \mathbb{Z}$ for $k = n-1$ and $0$ otherwise (we are still in the case $k \geq 2$). The reason of the difference for $k = n$ is still that the whole $\mathbb{R}^{n} \setminus \{0\}$ is a cycle only in Borel-Moore homology, and it turns out that it is non-trivial. For $k = n-1$, a non-trivial cycle for ordinary homology is the sphere $S^{n-1}$ embedded in $\mathbb{R}^{n} \setminus \{0\}$, but it becomes trivial in Borel-Moore homology since it is the boundary of the chain made by the disk without the origin $D^{n} \setminus \{0\}$, which is closed in $\mathbb{R}^{n} \setminus \{0\}$ but it is not compact, thus it is a chain only in Borel-Moore homology.

We now look at the remaining cases $k = 1$ and $k = 0$. For $k = 1$ the sequence becomes:
	\[\xymatrix{
	\cdots \ar[r] & 0 \ar[r] & 0 \ar[r] & H_{1}(S^{n}, \{N, S\}) \ar[r]^{\qquad \alpha} & \mathbb{Z} \oplus \mathbb{Z} \ar[r]^{\quad \beta} & \mathbb{Z} \ar[r] & \cdots
}\]
where the map $\beta$ is given by $\beta(n, m) = n-m$. Thus, $H_{1}^{BM}(\mathbb{R}^{n} \setminus \{0\}, \mathbb{Z}) \simeq \IIm\,\alpha = \Ker\,\beta \simeq \mathbb{Z}$. In ordinary homology $H_{1}(\mathbb{R}^{n} \setminus \{0\}, \mathbb{Z}) = 0$: the non-trivial Borel-Moore cycle is an open half-line from the origin to infinity. Finally, for $k = 0$ the sequence is:
	\[\xymatrix{
	\cdots \ar[r] & \mathbb{Z} \oplus \mathbb{Z} \ar[r]^{\quad \beta} & \mathbb{Z} \ar[r]^{\gamma \qquad\quad} & H_{0}(S^{n}, \{N, S\}) \ar[r] & 0 \ar[r] & \cdots
}\]
so that $H_{0}^{BM}(\mathbb{R}^{n} \setminus \{0\}, \mathbb{Z}) = \IIm\,\gamma$, but $\Ker\,\gamma = \IIm\,\beta = \mathbb{Z}$ so that $\gamma = 0$ thus $H_{0}^{BM}(\mathbb{R}^{n} \setminus \{0\}, \mathbb{Z}) = 0$. In ordinary homology $H_{0}(\mathbb{R}^{n} \setminus \{0\}, \mathbb{Z}) = \mathbb{Z}$: the difference is due to the fact that a point, which is non-trivial in ordinary homology, becomes the boundary of the Borel-Moore cycle made by half a line from it to infinity or from it to the origin.

\paragraph{}We now make an important remark. As one can see from the previous examples, \emph{Borel-Moore homology is not invariant under homotopy}, thus it is not an homology theory in the sense of Eilenberg and Steenrod (see \cite{ES}). It is invariant under homeomorphism, as one can see from the definition, thus it is a well-defined invariant of a topological space, but not up to homotopy. That's why it is less studied in the mathematical literature; however, as we will see soon, it naturally arises from Poincar\'e duality on manifolds.

\subsection{Borel-Moore cohomology}

To define ordinary singular cohomology, we define the set of \emph{$n$-cochains} of $X$ as:
	\[C^{n}(X, \mathbb{Z)} := \Hom(C_{n}(X, \mathbb{Z}), \mathbb{Z}) \simeq \prod_{\{\sigma^{n}: \Delta^{n} \rightarrow X\}} \mathbb{Z}
\]
(see appendix \ref{DirectSumProd}) and we define a coboundary operator $\delta^{n}: C^{n}(X, \mathbb{Z}) \rightarrow C^{n+1}(X, \mathbb{Z})$ given by:
	\[(\delta^{n}\varphi)(x) := \varphi(\partial_{n}x) \; .
\]
We thus define the \emph{singular cohomology groups} of $X$ as:
	\[H^{n}(X, \mathbb{Z}) := \Ker \,\delta^{n} \,/\, \IIm \,\delta^{n-1} \; .
\]
We can do the same for Borel-Moore version, and the cohomology we obtain, under suitable hypotesis that we now state, is the well-known \emph{cohomology with compact support}, i.e. the cohomology obtained restricting the boundary operator to cochains $\varphi$ such that there exists a compact subset $K_{\varphi} \subset X$ such that $\varphi$ is zero an all chains with image in $X \setminus K$. We call the associated cohomology groups $H^{n}_{\cpt}(X, \mathbb{Z})$. The hypoteses we need are that $X$ is Hausdorff and that there exists a countable family of compact sets $\{K_{n}\}_{n \in \mathbb{N}}$ such that $K_{n} \subset \Int(K_{n+1})$ and $\bigcup_{n \in \mathbb{N}} K_{n} = X$. They are always satisfied if $X$ is a manifold.

To prove that compactly-supported cohomology coincides with Borel-Moore cohomology, let us consider a Borel-Moore chain $\prod_{\sigma^{n}} k_{\sigma^{n}}$ and a cochain $\varphi$ with compact support $K_{\varphi}$. Then, for every point of $K_{\varphi}$ we choose a neighborhood realizing the definiton of local finteness and, by compactness, we exctract a finite subcover of $K_{\varphi}$: in this way we find a neighborhood of $K_{\varphi}$ intersecting finitely many simplices $\sigma^{n}$ with non-zero coefficient, thus $\varphi\bigl(\prod_{\sigma^{n}} k_{\sigma^{n}}\bigr)$ is well-defined. Viceversa, let us suppose that a cochain $\varphi$ is well-defined on every Borel-Moore chain and has not compact support. Let us consider a countable family of compact sets $\{K_{n}\}_{n \in \mathbb{N}}$ such that $K_{n} \subset \Int(K_{n+1})$ and $\bigcup_{n \in \mathbb{N}} K_{n} = X$. Then, we fix an $n$-simplex $\sigma^{n}_{1}$ such that $\varphi(\sigma^{n}_{1}) \neq 0$: up to change its sign, we can suppose that $\varphi(\sigma^{n}_{1}) > 0$. There exists $n_{1}$ such that $\IIm \sigma^{n}_{1} \subset K_{n_{1}}$. Then, since $\varphi$ has not compact support, we can fined another simplex $\sigma^{n}_{2}$ whose image is contained in $X \setminus K_{n_{1}}$ such that $\varphi(\sigma^{n}_{2}) > 0$. Keeping on in this way, we find infinitely many disjoint simplices $\{\sigma^{n}_{k}\}_{k \in \mathbb{N}}$ such that $\varphi(\sigma^{n}_{k}) > 0$ and $\IIm \varphi(\sigma^{n}_{k}) \subset (K_{n_{k}} \setminus K_{n_{k-1}})$. Being them disjoint $\prod_{k} \sigma^{n}_{k}$ is locally finite; we now prove that it is closed. Let us fix $x$ in the complement of the support: there exists $k$ such that $x \in \Int(K_{n_{k}}) \setminus K_{n_{k-2}}$, and the latter is open. In $\Int(K_{n_{k}}) \setminus K_{n_{k-2}}$ there are two simplices, so that their image is closed (since it is compact and $X$ is Hausdorff), so there exists a neighborhood of $x$ contained in the complement. Hence the complement is open so that the support is closed. Therefore $\prod_{k} \sigma^{n}_{k}$ is a Borel-Moore cycle, but $\varphi$ has infinite value on it. That's why $\varphi$ must have compact support.

\paragraph{}For a general manifold, Poincar\'e duality links ordinary homology (whose chains have compact support) with cohomology with compact support, and Borel-Moore homology with ordinary cohomology: we can say that Poincar\'e duality respects the support. Thus, on manifolds, the Poincar\'e dual of a cohomology class is naturally a Borel-Moore homology class. That's why Borel-Moore homology naturally appears on manifolds.

\subsection{Modified versions of Borel-Moore homology and cohomology}

We can introduce a suitable variation of Borel-Moore homology and compactly supported cohomology, which can be useful to describe D-brane charges. Let us consider a triple $(X, Y, r)$ where $X$ is a manifold, $Y \subset X$ a submanifold and $r: X \rightarrow Y$ a retraction (i.e. a surjective continuous map such that $r(y) = y \,\forall y \in Y$). We want to define a homology whose cycles are ``compact along $Y$ via $r$''. We thus give the following definition:
\begin{Def} $ $
\begin{itemize}
	\item A \emph{$(X, Y, r)$-Borel-Moore $n$-chain} is a generalized $n$-chain on $X$ which is \emph{locally finite}, has \emph{closed support} and is such that the image of its support via $r$ has \emph{compact closure} in $Y$.
	\item Calling $\partial_{n}^{BM (Y, r)}$ the boundary operator extended to locally finite generalized $n$-chains and restricted to $(X, Y, r)$-Borel-Moore ones, we define the \emph{$(X, Y, r)$-Borel-Moore singular homology groups} as:
		\[H_{n}^{BM}(X, Y, r, \mathbb{Z}) := \Ker \,\partial_{n}^{BM (Y, r)} \,/\, \IIm \,\partial_{n+1}^{BM (Y, r)} \; .
\]
\end{itemize}
\end{Def}
One particular case, which will be the interesting one for D-branes, is the one in which there exists a manifold $Z$ such that $X = Z \times Y$ and $r(z, y) = y$, i.e. $r$ is the natural projection. In this case, since we consider cycles which are compact on $Y$, they can go at infinity only along $Z$, that's why we have a canonical isomorphism:
	\[H_{n}^{BM}(Z \times Y, Y, \pi_{Y}, \mathbb{Z}) \simeq H_{n}((Z^{+} \times Y, \{\infty\} \times Y), \mathbb{Z})
\]
or, for a generic compactification $\overline{Z}$ of $Z$, we have $H_{n}^{BM}(Z \times Y, Y, \pi_{Y}, \mathbb{Z}) \simeq H_{n}((\overline{Z} \times Y, (\overline{Z} \times Y) \setminus (Z \times Y)), \mathbb{Z})$.

\paragraph{}Let us consider the example of $\mathbb{R}^{n} = \mathbb{R}^{m} \times \mathbb{R}^{n-m}$. In this case we have $H_{k}^{BM}(\mathbb{R}^{n}, \mathbb{R}^{n-m},$ $\pi_{n-m}, \mathbb{Z}) \simeq H_{n}((S^{m} \times \mathbb{R}^{n-m}, \{N\} \times \mathbb{R}^{n-m}), \mathbb{Z}) \simeq \tilde{H}_{k}((S^{m} \times \mathbb{R}^{n-m})/(\{N\} \times \mathbb{R}^{n-m}), \mathbb{Z})$, but since the latter space retracts on $S^{m}$ we obtain $\mathbb{Z}$ for $k = m$ and $0$ otherwise. For ordinary homology we would have $\mathbb{Z}$ for $k = 0$ and $0$ otherwise, while for Borel-Moore homology we would have $\mathbb{Z}$ for $k = n$ and $0$ otherwise. The reason is that, for $k = n$, the whole $\mathbb{R}^{n}$ is a non-trivial Borel-Moore cycle, but it is not a cycle in the modified version (for $m < n$) since it is non-compact also in the last $(n-m)$-directions. For $k = m$, one non-trivial cycle in the modified Borel-Moore homology is $\mathbb{R}^{m} \times \{0\}$, which is not a cycle in ordinary homology since it is non-compact, and which is trivial in standard Borel-Moore homology since it is the boundary of $\{(v, w) \in \mathbb{R}^{m} \times \mathbb{R}^{n-m}: v_{i} \geq 0 \, \forall i = 1, \ldots,m\}$; the latter is not a chain in modified Borel-Moore homology since it non-compact also in the last $(n-m)$ directions, thus it does not make the cycle $\mathbb{R}^{m} \times \{0\}$ trivial in this case. For $k = 0$, the origin, which is a non trivial cycle in ordinary homology, becomes trivial also in the modified Borel-Moore homology: it is enough to take a half-line going to infinity along the first $k$ directions, e.g. on the first $k$ coordinate half-axes.

\paragraph{}\textbf{Remark:} We must ask that the projection has compact closure since in general is not closed. For example in $\mathbb{R} \times \mathbb{R}$ the graph of the function $y = \tan(x)$ for $x \in (-\frac{\pi}{2}, \frac{\pi}{2})$ has open projection on the first factor. However, since the closure of a set contains its boundary points, that fact of having compact closure is the right translation of the idea of not to go to infinity along $Y$.

\paragraph{}The cohomological version of this modified theory is defined analogously and it coincides with the cohomology which compact support along $Y$ via $r$. The proof is the same considered for the general case. In particular, Poincar\'e duality gives an isomorphism between modified Borel-Moore homology and cohomology.

\subsection{Borel-Moore homology and currents}

Since Borel-Moore homology is isomorphic to ordinary cohomology via Poincar\'e duality, it is also isomorphic to the cohomology of currents, and the same for the modified version. We analyze this isomorphism in more detail. We recall (see \cite{GH}) that, for $X$ an $n$-dimensional manifold, there are two isomorphisms:
	\[\begin{array}{ll}
	\varphi_{1}: &H^{k}_{dR}(X) \overset{\simeq}\longrightarrow H^{k}_{crn}(X)\\
	&[\omega] \longrightarrow [T_{\omega}] \\ \\
	\varphi_{2}: &H_{n-k}(X, \mathbb{R}) \overset{\simeq}\longrightarrow H^{k}_{crn,cpt}(X)\\
	&[\Gamma] \longrightarrow [\delta(\Gamma)]
\end{array}\]
where $T_{\omega}(\varphi) := \int_{X}(\omega \wedge \varphi)$ and $\delta(\Gamma)(\varphi) := \int_{\Gamma}\varphi$ for $\varphi$ compactly-supported $(n-k)$-form. It is easy to verify that $\varphi_{1}^{-1} \circ \varphi_{2}: H_{n-k}(X, \mathbb{R}) \longrightarrow H^{k}_{dR,cpt}(X)$ is exactly Poincar\'e duality. The previous isomorphisms mean that currents encodes both homology and cohomology: for example, a $\delta$-current supported over a cycle can be identified both with its support, which is a homology cycle, or with an approximating sequence of bump forms picked over such a supports, which are all cohomologous and determine the Poincar\'e dual of the support.

Of course a $\delta$-current can be picked also over a non-compact cycle: the test form $\varphi$ is compactly-supported by definition, hence the integral is well-defined. That's why currents are more naturally associated to Borel-Moore cycles, i.e. we can extend $\varphi_{2}$ to:
	\[\begin{array}{ll}
	\varphi_{2}^{BM}: &H_{n-k}^{BM}(X, \mathbb{R}) \overset{\simeq}\longrightarrow H^{k}_{crn}(X)\\
	&[\Gamma] \longrightarrow [\delta(\Gamma)]
\end{array}\]
and the fact that this is an isomorphism means that every current is cohomologus to a $\delta$-current over a Borel-Moore cycle. The isomorphism $\varphi_{2}^{BM}$ can be defined without problems for the modified versions, assuming in both the l.h.s. and the r.h.s. the suitable compactness hypotesis.

\section{Borel-Moore homology and D-branes}

Now that we are able to deal with non-compact homology cycles, we can consider branes with are not necessarily particles in the non-compact space-time directions and also write the charge equations considering the whole world-volume.

\subsection{Linear and general D-branes}

We recall that the space-time manifold is $S = \mathbb{R} \times M$ with $M = \mathbb{R}^{3} \times X$ for $X$ a 6-dimensional compact manifold. We consider for the moment D-branes which are lines or planes in the non-compact space direction. If $Y_{p,t}$ is the volume at time $t$, we call $V = \pi_{\mathbb{R}^{3}}(Y_{p,t})$ (one simple case is $Y_{p,t} = V \times Y'_{p-k,t}$ with $Y'_{p-k,t} \subset X$, but this is not necessary). In this case, to define their charge we use modified Borel-Moore homology, considering that their volume is compact in the directions $V^{\bot} \times X$. We thus write the charge equations as:
\begin{equation}\label{ChargeBM}
	\left\{ \begin{array}{l}
		[\, d_{M}G^{s}_{8-p} \,] = \PD_{BM(M, V^{\bot}, \pi)}(q \cdot Y_{p,t}) \\
		d*_{9}G^{s}_{8-p} = 0 \; .
	\end{array} \right.
\end{equation}
We have chosen the couple $(M, V^{\bot})$ but it is equivalent to the couple $(M, V^{\bot} \times X)$, since $X$ is compact. We can solve equations \eqref{ChargeBM} in a way analogue to the particle case. We obtain up to isomorphism $H_{\cpt}^{p}(V^{\bot} \times X) \simeq H^{p-3+k}(X)$, the isomorphism being given by $3-k$ applications of \eqref{IsoCptCohom}. Thus, instead of solving
	\[dG^{s}_{8-p} = e_{1}(x)\,dx_{1} \wedge e_{2}(x)\,dx_{2} \wedge e_{3}(x)\,dx_{3} \wedge \alpha + d\xi_{cpt}
\]
as in the ordinary case, we have to solve one of the two equations:
	\[dG^{s}_{8-p} = e_{1}(x)\,dx_{1} \wedge \alpha + d\xi_{cpt} \qquad dG^{s}_{8-p} = e_{1}(x)\,dx_{1} \wedge e_{2}(x)\,dx_{2} \wedge \alpha + d\xi_{cpt}
\]
depending whether $k = 2$ or $k = 1$. Then, the same procedure considered before applies.

\paragraph{}As ordinary homology is homotopy-invariant, similary the modified Borel-Moore homology of a split-manifold $A \times B$ is invariant under homotopies involving only $B$, i.e. under homotopies of the form $F_{t}(a,b) = (a, F'_{t}(b))$. In particular, the modified Borel-Moore homology of $V \times V^{\bot} \times X$, non-compact only on $V$, is isomorphic to the one of $V \times X$ since $V^{\bot}$ retracts to a point. Now the only non-compact directions are the one on which cycles are allowed to be non-compact, thus we reduce to standard Borel-Moore homology. The situation is reversed for cohomology, since the ordinary one has \emph{non}-compact support in general, and that's the one which is homotopy-invariant. Thus, the modified Borel-Moore cohomology of $V \times V^{\bot} \times X$, which is the cohomology with compact support on $V^{\bot}$, is isomorphic to the one of $V^{\bot} \times X$. Since we are left only with compact directions, we reduce to the usual compactly-supported cohomology.

\paragraph{Remarks:}
\begin{itemize}
	\item One might think that we can always use standard Borel-Moore homology, since, having no hypoteses on the compactness of the cycles, it includes any kind of D-brane. This is not correct. In fact, let us consider a particle brane with worldvolume $\mathbb{R} \times Y_{p}$ in a fixed reference frame. Then, if we consider it as a Borel-Moore cycle, it is the boundary of $\mathbb{R} \times H^{3} \times Y_{p}$ for $H^{3} = \{(x,y,z): x,y,z \geq 0\}$, thus it has no charge. In general, if we want a cycle to be non-trivial, we must assume the necessary compactness hypoteses.
	\item We considered only lines or planes and not generic curves or surfaces. In the latter case, if we do not assume that they go at infinity along a fixed plane of the same dimension, we must consider only the direction at infinity of the brane itself, thus we should consider $H_{p}((S \cup (WY_{p})^{+}, \{\infty\}), \mathbb{Z})$ or in general $H_{p}((S \cup \overline{WY_{p}}, (S \cup \overline{WY_{p}}) \setminus (S \cup WY_{p})), \mathbb{Z})$. In terms of cycles in $S$ we must ask that their closure on $\overline{S}$ intersects $\partial S$ only on $\overline{WY_{p}}$. This is less natural but it works without any hypoteses.
\end{itemize}

\subsection{D-brane charge and world-volume}

Using modified Borel-Moore homology we can describe D-brane charges directly from the world-volume, without restricting to a fixed instant of time. In particular, let us consider a brane which is a particle in the non-compact space-time directions. In a fixed reference frame in which it is fixed in the origin we can rewrite the charge equation $(dG_{8-p})\vert_{M_{t}} = \PD_{M_{t}}(q \cdot Y_{p,t})$ as $[dG_{8-p}] = \PD_{BM(S, \mathbb{R}^{3}, \pi)}(q \cdot WY_{p})$. However, with Borel-Moore homology, we can write a Lorentz-invariant expression holding for every brane regardless of the behaviour of the brane in the $\mathbb{R}^{1,3}$-directions. We call $V = \pi_{\mathbb{R}^{1,3}}(WY_{p})$ and we get:
\begin{equation}\label{ChargeWWGeneric}
	[dG_{8-p}] = \PD_{BM(S, V^{\bot}, \pi)}(q \cdot WY_{p}) \; .
\end{equation}
There are no problems in considering always Borel-Moore homology in time-direction since any stable world-volume is non-trivial in that direction. If we would want to define an instanton charge, for a world-volume $\{t\} \times Y_{p}$, we could use analogue equations with ordinary $\PD$ (i.e. from ordinary homology to compactly supported cohomology in $\mathbb{R}^{1,3} \times X$); of course this is not a charge conserved in time, it is a trajectory charge but computed at a fixed instant since the trajectory itself is at a fixed instant.

Equation \eqref{ChargeWWGeneric} applies also to classical electromagnetism theory, since we can write $[d*F] = \PD_{BM(\mathbb{R}^{1,3}, w^{\bot}, \pi)}(q \cdot w)$.

\subsection{Space-filling D-branes}

Up to now we have not considered the most common D-branes, i.e. the space-filling ones. That's because, in this setting, their total charge would be zero. In fact, we should consider Borel-Moore homology which is non-compact in all $\mathbb{R}^{3}$, but in this case the Poincar\'e dual gives an ordinary cohomology class, thus, if is equal to $dG_{8-p}$, it is necessarily the trivial class, so the charge equations have no solution. The physical reason is that there are no directions at infinity where the flux could go, so that there is no charge, as it happens for an electron on a compact space. For an electron, we should put an anti-electron in another point of the space, so that fluxes go from one to the other. For D-branes we have two main possibilities:
\begin{itemize}
	\item There is an anti-brane, so that the total charge is $0$. In this case, we could imagine to compute each of the two opposite charges: to do this, if $Y$ is the brane and $\overline{Y}$ the anti-brane, we should solve the equations for Ramond-Ramond field in $(\mathbb{R}^{1,3} \times X) \setminus W\overline{Y}_{p}$ for $Y_{p}$ and in $(\mathbb{R}^{1,3} \times X) \setminus WY_{p}$ for the anti-brane. The result should give Ramond-Ramond fields extendable on all $(\mathbb{R}^{1,3} \times X)$ and closed, since the Poincar\'e dual is the zero class.
	\item There is an orientifold plane $O$ assorbing fluxes, so that we compute Ramond-Ramond fields in $(\mathbb{R}^{1,3} \times X) \setminus O$. 
\end{itemize}
In any case, we must consider a manifold which is not of the form $\mathbb{R}^{1,3} \times X$ with $X$ compact. However we can use without problems equation \eqref{ChargeWWGeneric}. In fact, an orientifold or an anti-brane is of the form $O = \mathbb{R}^{1,3} \times O'$, so we consider $(\mathbb{R}^{1,3} \times X \setminus O')$ and we reduce to the previous case with the only difference that the internal manifold is non-compact. Here we must consider the cohomology $BM(S, \mathbb{R}^{3-k} \times (X \setminus O'), \pi)$ and not $BM(S, \mathbb{R}^{3-k}, \pi)$ since the brane must be far from the orientifold in type II superstring theory.

\section{Wess-Zumino action}\label{WessZumino}

\subsection{Definition of the action}

If we consider a small charge $q$ moving in an electromagnetic field, the action of the particle minimally couples to the electromagnetic field via the potential, i.e. we add the term $q \int_{\gamma} A$. Such an integral is actually the holonomy of the line bundle over the curve $\gamma$, and in a general background the field strength $F$ can be topologically non-trivial, so $A$ is locally defined and has gauge transformations. The problem is that, as explained in \cite{BFS}, holonomy is a well-defined function on closed curves, while it is a section of a line bundle over the space of open curves. However, at classical level, when we minimize the action we do it for curves connecting two fixed points (they can be at infinity, in case the bundle extends to the closure $\overline{S}$). In this case, if we fix a trivialization of the bundle near the two points we define holonomy as a number; actually, on a connected component of curves likining $x_{1}$ and $x_{2}$ and homotopic one to the other, we can choose a trivialization along all the curves and this is equivalent to fixing a potential $A$ . In this case, if we change the potential by a gauge transformation $A \rightarrow A + \Phi$, then $S'(\gamma) = S(\gamma) + \int_{\gamma} \Phi$, but the summand $\int_{\gamma} \Phi$ is independent on $\gamma$ since $\int_{\gamma} \Phi - \int_{\gamma'} \Phi = \int_{\gamma - \gamma'}\Phi = 0$ being $\Phi$ closed and $\gamma - \gamma'$ contractible. We can have different constants $\int_{\gamma} \Phi$ on each connected component of the space of open curves between $x_{1}$ and $x_{2}$, but this has no influence on the minima or in general on stationary points. At quantum level, since the Wilson loop is an observable, on our background we have a fixed holonomy for the connection, then we must also consider the case in which $\gamma - \gamma'$ is a non-trivial cycle: in this case the difference is the Wilson loop of a \emph{geometrically trivial} connection over $\gamma - \gamma'$, which is quantized for bundles, i.e. for $F$ quantized, thus the holonomy is zero at the exponential, i.e. for the partition function. For the D-brane the same considerations hold, the minimal coupling being the Wess-Zumino action:
	\[S_{WZ} = \int_{WY_{p}} C_{p+1} \; .
\]
We must assume that $G_{p+2}$ is closed and quantized, i.e. that it represents an integral cohomology class. We see it as the curvature of a $p$-gerbe on $S$, and we assume that this gerbe is endowed with a connection $C_{p+1}$, so that $dC_{p+1} = G_{p+2}$. Actually the local forms $C_{p+1}$ are just the top forms representing the gerbe connection: a complete connection is given by a set of local forms from the degree $p+1$ to degree one, to end with transition functions $g_{\alpha_{0}\cdots \alpha_{p+1}}$, as explained in appendix \ref{AppGerbes}. In particular, for $\Omega^{p}_{\mathbb{R}}$ the sheaf of smooth $p$-forms on $S$ and $\underline{S}^{1}$ the sheaf of $S^{1}$-valued smooth functions on $S$, the background data is a gerbe with connection:
	\[\mathcal{G}_{p} \in \check{H}^{p+1}(S, \underline{S}^{1} \rightarrow \Omega^{1}_{\mathbb{R}} \rightarrow \cdots \rightarrow \Omega^{p+1}_{\mathbb{R}})
\]
whose curvature is $G_{p+2}$ and whose holonomy on the corresponding world-volumes are the Wess-Zumino actions. For a brief review about the holonomy of gerbes we refer to \cite{BFS} chap. 3 and references therein: that discussion can be immediately generalized to $p$-gerbes, considering triangulations of dimension $p+1$ instead of $2$. In particular, in the definition of the holonomy we must consider all the intermediate forms defining the connection, a $k$-form being integrated on the $k$-faces of the triangulation of $WY_{p}$. The top forms $C_{p+1}$ are only a small piece of information, so that the notation $\int_{WY_{p}} C_{p+1}$ is actually approximate.

\subsection{Holonomy and boundary conditions}

As for line bundles, the holonomy of a $p$-gerbe is well-defined only on \emph{closed} $(p+1)$-manifolds, i.e. on manifolds without boundary, while $WY_{p}$, being the classical trajectory of the D-brane, is in general defined for all times so that it has a boundary at the limit time-coordinates $-\infty$ and $+\infty$, or equivalently it has a boundary contained in the boundary of $S$. As for line bundles, if we fix boundary conditions we have no problems with the partition function. In general, the path-integral gives a section of a bundle, so there are no problems. If we want to define the holonomy as a number, we must give boundary conditions at infinity, but since the forms $C_{p+1}$ are defined only locally and only up to gauge transformations, it is not immediately clear how to impose boundary conditions on them at infinity.

We refer to \cite{BFS} chap. 4 for a discussion about $\rm\check{C}$ech hypercohomology and trivializations of gerbes. We can generalize the discussion there to $p$-gerbes. In particular, we consider the compactification $\overline{S}$ of space-time making such that $\overline{S}$ is a manifold with boundary and $S$ its interior, so that the infinity of $S$ becomes the boundary $\partial \overline{S}$. For example, for $\mathbb{R}^{1,3} \times X$ such a compactification is $D^{4} \times X$ for $D^{4}$ the $4$-disc, so that the boundary is $S^{4} \times X$. Generalizing the explanation in \cite{BFS} to $p$-gerbes we can define the holonomy of a $p$-gerbe $\mathcal{G}_{p}$ on $S$ along a $(p+1)$-submanifold with boundary. This holonomy is not a function, it is a section of a line bundle over the space of maps from open $(p+1)$-manifolds to $S$. In particular, if we fix a $(p+1)$-manifold $\Sigma$ and we endow the space $\Maps(\Sigma, S)$ with a suitable topology, the holonomy of $\mathcal{G}_{p}$ is a section of a line bundle over $\Maps(\Sigma, S)$. However, if we fix a subspace $T \subset S$ such that $\mathcal{G}_{p}\vert_{T}$ is trivial, and we consider only maps such that $\varphi(\partial \Sigma) \subset T$, then the line bundle becomes trivial. This is not enough to have a well-defined holonomy, since we do not have a preferred trivialization. However, a trivialization of $\mathcal{G}_{p}\vert_{T}$ determines canonically a trivialization of the line bundle, making the holonomy a well-defined function. In this case, we consider $\partial S$ as the subset $T$ on which the gerbe must be trivial, since the boundary of the compactified world-volume $WY_{p}$ lies in the boundary of $\overline{S}$. Thus, the background data must not only be a gerbe with connection $\mathcal{G}_{p}$ on $\overline{S}$ which is trivial on $\partial S$, but also a fixed trivialization of it. The $\rm\check{C}$ech double-complex to consider is then:
\[\xymatrix{
	\check{C}^{0}(\overline{S}, \Omega^{p+1}_{\mathbb{R}}) \oplus \check{C}^{0}(\partial S, \Omega^{p}_{\mathbb{R}}) \ar[r]^{\check{\delta}^{0}} & \check{C}^{1}(\overline{S}, \Omega^{p+1}_{\mathbb{R}}) \oplus \check{C}^{1}(\partial S, \Omega^{p}_{\mathbb{R}}) \ar[r]^{\check{\delta}^{1}} & \check{C}^{2}(\overline{S}, \Omega^{p+1}_{\mathbb{R}}) \oplus \check{C}^{2}(\partial S, \Omega^{p}_{\mathbb{R}}) \ar[r]^{\phantom{XXXXXXXI}\check{\delta}^{2}} & \cdots \\
	\qquad\vdots\qquad \ar[r]^{\check{\delta}^{0}} \ar[u]^{d} & \qquad\vdots\qquad \ar[r]^{\check{\delta}^{1}} \ar[u]^{d} & \qquad\vdots\qquad \ar[r]^{\phantom{XXX}\check{\delta}^{2}} \ar[u]^{d} & \cdots \\
	\check{C}^{0}(\overline{S}, \Omega^{1}_{\mathbb{R}}) \oplus \check{C}^{0}(\partial S, \underline{S}^{1}) \ar[r]^{\check{\delta}^{0}} \ar[u]^{d} & \check{C}^{1}(\overline{S}, \Omega^{1}_{\mathbb{R}}) \oplus \check{C}^{1}(\partial S, \underline{S}^{1}) \ar[r]^{\check{\delta}^{1}} \ar[u]^{d} & \check{C}^{2}(\overline{S}, \Omega^{1}_{\mathbb{R}}) \oplus \check{C}^{2}(\partial S, \underline{S}^{1}) \ar[r]^{\phantom{XXXXXXX}\check{\delta}^{2}} \ar[u]^{d} & \cdots \\
	\check{C}^{0}(\overline{S}, \underline{S}^{1}) \ar[r]^{\check{\delta}^{0}} \ar[u]^{\tilde{d}} & \check{C}^{1}(\overline{S}, \underline{S}^{1}) \ar[r]^{\check{\delta}^{1}} \ar[u]^{\tilde{d}} & \check{C}^{2}(\overline{S}, \underline{S}^{1}) \ar[r]^{\phantom{XXX}\check{\delta}^{2}} \ar[u]^{\tilde{d}} & \cdots
	}
\]
and we denote by $\check{H}^{\bullet}(\overline{S}, \underline{S}^{1} \rightarrow \Omega^{1}_{\mathbb{R}} \rightarrow \cdots \rightarrow \Omega^{p+1}_{\mathbb{R}}, \partial S)$ the hypercohomology of this complex. Thus, if we want to give boundary conditions to field in order to make Wess-Zumino action a well-defined number for a fixed trajectory extending in time from $-\infty$ to $+\infty$, we must give as background data a gerbe with trivialization:
	\[\mathcal{G}_{p} \in \check{H}^{p+1}(\overline{S}, \underline{S}^{1} \rightarrow \Omega^{1}_{\mathbb{R}} \rightarrow \cdots \rightarrow \Omega^{p+1}_{\mathbb{R}}, \partial S) \; .
\]

\section{Remarks about the $*$-problem}

\subsection{$*$-problem}

In a generic configuration we can have both electric and magnetic sources:
\[\begin{array}{lll}
	dG_{8-p} = q \cdot \delta(WY_{p}) & & dG_{p+2} = q' \cdot \delta(WY_{6-p}) \\
	q = \frac{1}{l} \int_{L} G_{8-p} & & q' = \frac{1}{l'} \int_{L'} G_{p+2}
\end{array}\]
from which we obtain the cohomological relations:
	\[\begin{array}{lll}
	[\, dG_{8-p} \,] = \PD_{BM(S, V^{\bot}, \pi)}(q \cdot WY_{p,t}) & & [\, dG_{p+2} \,] = \PD_{BM(S, V'^{\bot}, \pi)}(q' \cdot WY_{6-p,t}) \; .
\end{array}\]
For Dirac quantization condition, if both the branes $Y_{p+2}$ and $Y_{6-p}$ are present, then $G_{8-p}$ and $G_{p+2}$ must be both quantized, but this is not compatible with the relation $G_{8-p} = *G_{p+2}$, since the Hodge dual of a quantized form is in general not quantized. This can be seen from the fact that, varying the metric, the coefficients of the Hodge dual of a fixed form change continuosly, thus they give a non-quantized form for a generic metric. This presents a serious problem, say, for $G_{5}$ which is self-dual and so not quantizable. This problem is well-known and there are various proposals on how to solve it. One is to quantize, for IIA theory, only $G_{p}$ for $p < 5$, but this creates some problems (see \cite{MW} and references therein), and does not solve the problem of $G_{5}$ for IIB theory. We can actually find backgrounds in which we can avoid this problem, as suggested in \cite{Evslin}, supposing that the time $\mathbb{R}$ and the space $M$ are orthogonal with respect to the background metric. In this case, we can divide the Ramond-Ramond fields in spatial and temporal part, the temporal part being made by the summands containing a temporal leg $dt \wedge$. In this case, thanks to the orthogonality, the Hodge-$*$ exchanges temporal and spatial part, giving a separation between electric and magnetic fluxes. Thus, we see the D-branes as magnetic sources, and we quantize only the spatial part avoiding the $*$-problem. Of course this is not a Lorentz-invariant formulation, since we must select some preferred reference frames in which the fields are static, but it gives us a model in which we can correctly define all the data involved.

\subsection{A non Lorentz-invariant solution}

We recall the decomposition \eqref{RRSplit}:
	\[G_{p} = G^{s}_{p} + dt \wedge G^{t}_{p-1}
\]
where $G^{s}_{p}$ and $G^{t}_{p-1}$ have no temporal legs. We suppose the coefficients time-independent, so that these forms are \emph{spatial}, i.e. they restrict to well-defined forms on the space $M = \mathbb{R}^{3} \times X$. Thus, calling $d_{9}$ the exterior differential in $M$:
	\[dG_{p} = d_{9}G^{s}_{p} - dt \wedge d_{9}G^{t}_{p-1}
\]
and, in particular, $dG_{p} = 0$ if and only if $d_{9}G^{s}_{p} = 0$ and $d_{9}G^{t}_{p-1} = 0$ (this is what happens when no D-branes are present). 

It is now easy to describe self-duality condition of the polyform $G$, i.e. $G = *G$. Actually, since $*^{2}\vert_{\Lambda^{p}T^{*}(\mathbb{R}^{1,3} \times X)} = -(-1)^{p(10-p)}$ (see appendix \ref{HodgeMinkowski}), this condition is consistent for $p$ odd, i.e. for IIB theory, while for IIA theory we must fix $G_{p}$ for $p < 5$ and impose $G_{5+q} = *G_{5-q}$, so that $G_{5-q} = -*G_{5+q}$ (otherwise we should ask $*G = iG$, but they are real fields). We now consider IIB theory to simplify the notation. Self-duality condition becomes by equation \eqref{RRHodge} for $p$ odd (so that $(-1)^{p} = -1$):
\begin{equation}\label{RRHodgeSplit}
	\left\{ \begin{array}{lll} G^{s}_{10-p} & = & -*_{9}G^{t}_{p-1} \\ G^{t}_{(10-p)-1} & = & -*_{9}G^{s}_{p} \end{array} \right.
\end{equation}
and, since $*_{9}$ is Euclidean and $9$ is odd so that $*_{9}^{2} = 1$, the two equations are equivalent. Thus, in this picture, self-duality condition means that the temporal part is complitely determined by the spatial part via $-*_{9}$ and viceversa.

\paragraph{}Let us now consider D-branes $Y_{p}$ which are \emph{particles} with respect to the non-compact dimensions $\mathbb{R}^{1,3}$. We consider for simplicity $p \neq 3$ and $WY_{p} \cap WY_{6-p} = \emptyset$, otherwise we should introduce the corrections discussed in the previous section. We now consider the D-brane as \emph{magnetic} soruces and we suppose there are \emph{no electric sources}, so that we get, for fixed instant $t$:
	\[\left\{ \begin{array}{l}
	d_{M}G^{s}_{8-p} = \delta_{M}(q \cdot Y_{p,t}) \\
	d*_{9}G^{s}_{8-p} = 0
\end{array} \right. \qquad
\left\{ \begin{array}{l}
	d_{M}G_{p+2}^{s} = \delta_{M}(q \cdot Y_{6-p,t}) \\
	d*_{9}G^{s}_{p+2} = 0
\end{array} \right. \]
which becomes:
	\[\left\{ \begin{array}{l}
	d_{M}G^{s}_{8-p} = \delta_{M}(q \cdot Y_{p,t}) \\
	dG^{t}_{p+1} = 0
\end{array} \right. \qquad
\left\{ \begin{array}{l}
	d_{M}G_{p+2}^{s} = \delta_{M}(q \cdot Y_{6-p,t}) \\
	dG^{t}_{7-p} = 0
\end{array} \right. \]
so that they involve four independent forms and there is no inconsistency any more. The two systems are now completely independent, so that we can solve them as in section 2.

\section{Conclusions and perspectives}

We have seen the cohomological description of D-brane chages and Ramond-Ramond fields, which generalizes classical electromagnetism. As we said in the introduction, the analysis of this paper should be reproduced for the K-theoretical description of D-brane charges. In particular, discussing the geometrical meaning of Wess-Zumino action, we should also consider gauge and gravitational couplings. Moreover, in this paper we always supposed that the $H$-flux was vanishing. Without this assumption, ordinary cohomology should be replaced by twisted cohomology, as described in \cite{Kapustin}. However, we do not know a correct definition of integral twisted homology, thus it is hard to implement Dirac quantization condition in this setting. Maybe the proper setting is twisted K-theory, for which there is also the integral version (see \cite{AS}); we will discuss these problems in a future work.

\section*{Acknowledgements}

We would like to thank Loriano Bonora, Sergio Cecotti, Raffaele Savelli and Andrea Prudenziati for useful discussions.

\appendix

\section{p-Gerbes}\label{AppGerbes}

We refer to appendix B of \cite{BFS} and references therein for an introduction to gerbes. Here we just generalize the discussion to $p$-gerbes. In particular, we recall that a gerbe with connection is given by an element of the $\rm\check{C}$ech hypercohomology group:
	\[\check{H}^{2}(X, \underline{S}^{1} \rightarrow \Omega^{1}_{\mathbb{R}} \rightarrow \Omega^{2}_{\mathbb{R}}) \; .
\]
We thus define a $p$-gerbe with connection as an element of the $\rm\check{C}$ech hypercohomology group:
	\[\check{H}^{p+1}(X, \underline{S}^{1} \rightarrow \Omega^{1}_{\mathbb{R}} \rightarrow \cdots \rightarrow \Omega^{p+1}_{\mathbb{R}}) \; .
\]
The $\rm\check{C}$ech double complex with respect to a good cover $\mathfrak{U}$ is given by:
\[\xymatrix{
	\check{C}^{0}(\mathfrak{U}, \Omega^{p+1}_{\mathbb{R}}) \ar[r]^{\check{\delta}^{0}} & \check{C}^{1}(\mathfrak{U}, \Omega^{p+1}_{\mathbb{R}}) \ar[r]^{\check{\delta}^{1}} & \check{C}^{2}(\mathfrak{U}, \Omega^{p+1}_{\mathbb{R}})  \ar[r]^{\phantom{XXX}\check{\delta}^{2}} & \cdots \\
	\qquad\vdots\qquad \ar[r]^{\check{\delta}^{0}} \ar[u]^{d} & \qquad\vdots\qquad \ar[r]^{\check{\delta}^{1}} \ar[u]^{d} & \qquad\vdots\qquad \ar[r]^{\phantom{XXX}\check{\delta}^{2}} \ar[u]^{d} & \cdots \\
	\check{C}^{0}(\mathfrak{U}, \Omega^{1}_{\mathbb{R}}) \ar[r]^{\check{\delta}^{0}} \ar[u]^{d} & \check{C}^{1}(\mathfrak{U}, \Omega^{1}_{\mathbb{R}}) \ar[r]^{\check{\delta}^{1}} \ar[u]^{d} & \check{C}^{2}(\mathfrak{U}, \Omega^{1}_{\mathbb{R}}) \ar[r]^{\phantom{XXX}\check{\delta}^{2}} \ar[u]^{d} & \cdots \\
	\check{C}^{0}(\mathfrak{U}, \underline{S}^{1}) \ar[r]^{\check{\delta}^{0}} \ar[u]^{\tilde{d}} & \check{C}^{1}(\mathfrak{U}, \underline{S}^{1}) \ar[r]^{\check{\delta}^{1}} \ar[u]^{\tilde{d}} & \check{C}^{2}(\mathfrak{U}, \underline{S}^{1}) \ar[r]^{\phantom{XXX}\check{\delta}^{2}} \ar[u]^{\tilde{d}} & \cdots
	}
\]
so that $\check{C}^{p+1}(\mathfrak{U}, \underline{S}^{1} \rightarrow \Omega^{1}_{\mathbb{R}} \rightarrow \cdots \rightarrow \Omega^{p+1}_{\mathbb{R}}) = \check{C}^{p+1}(\mathfrak{U}, \underline{S}^{1}) \oplus \check{C}^{p}(\mathfrak{U}, \Omega^{1}_{\mathbb{R}}) \oplus \cdots \oplus \check{C}^{0}(\mathfrak{U}, \Omega^{p+1}_{\mathbb{R}})$. Thus, a representative hypercocycle of a gerbe with connection is a sequence $(g_{\alpha_{0} \cdots \alpha_{p+1}},$ $(C_{1})_{\alpha_{0} \cdots \alpha_{p}}, \ldots, (C_{p+1})_{\alpha_{0}})$, while summing an hypercoboundary represents a gauge transformation. It is easy to verify that for hypercocycles the local forms $dC_{p+1}$ glue to a global one $G_{p+2}$ which is the curvature of the gerbe. Thus, the data of the superstring background must be an equivalence class like this one, not only $C_{p+1}$.

Given a $p$-gerbe with connection $[(g_{\alpha_{0} \cdots \alpha_{p+1}}, (C_{1})_{\alpha_{0} \cdots \alpha_{p}}, \ldots, (C_{p+1})_{\alpha_{0}})]$, we can forget the connection and consider just the $p$-gerbe $\mathcal{G} = [g_{\alpha_{0} \cdots \alpha_{p+1}}] \in \check{H}^{p+1}(X, \underline{S}^{1})$. Then we can define the \emph{first Chern class} $c_{1}(\mathcal{G}) \in \check{H}^{p+2}(X, \mathbb{Z})$: we write the transition functions as $g_{\alpha_{0} \cdots \alpha_{p+1}} = e^{2\pi i \rho_{\alpha_{0} \cdots \alpha_{p+1}}}$ so that $\check{\delta}\{\rho_{\alpha_{0} \cdots \alpha_{p+1}}\} = \{c_{\alpha_{0} \cdots \alpha_{p+1} \alpha_{p+1}}\}$ with $c_{\alpha_{0} \cdots \alpha_{p+1} \alpha_{p+1}} \in \mathbb{Z}$. We then consider $c_{1}(\mathcal{G}) := [\{c_{\alpha_{0} \cdots \alpha_{p+1} \alpha_{p+1}}\}] \in \check{H}^{p+2}(X, \mathbb{Z})$. One can see that the de-Rham cohomology class of the curvature $G_{p+2}$ corresponds to $c_{1}(\mathcal{G}) \otimes_{\mathbb{Z}} \mathbb{R}$ under the canonical isomorphism between de-Rham cohomology and $\rm\check{C}$ech cohomology of the constant sheaf $\mathbb{R}$.

\section{Hodge-$*$ with Minkowskian signature}\label{HodgeMinkowski}

Let $V$ be an oriented vector space of dimension $n$ with a fixed \emph{Euclidean} metric. We recall that Hodge-$*$ operation is defined on the exterior algebra $\Lambda^{\bullet}V^{*}$ by:
\begin{equation}\label{HodgeStar}
	\alpha \wedge *\beta = \langle \alpha, \beta \rangle \cdot \vol
\end{equation}
where $\vol$ is the unit oriented volume form, given by $\vol = e_{1}^{*} \wedge \ldots \wedge e_{n}^{*}$ for $\{e_{1}, \ldots, e_{n}\}$ an oriented orthonormal basis. In particular, for $\alpha = e_{i_{1}}^{*} \wedge \ldots \wedge e_{i_{p}}^{*}$, equation \eqref{HodgeStar} with $\beta = \alpha$ gives $*\alpha = \varepsilon^{i_{1} \cdots i_{p} j_{1} \cdots j_{n-p}} e_{j_{1}}^{*} \wedge \ldots \wedge e_{j_{n-p}}^{*}$.

If the metric is Minkowskian, definition of Hodge-$*$ via \eqref{HodgeStar} still holds. Moreover, the volume form is the same, i.e. $\vol = e_{0}^{*} \wedge \ldots \wedge e_{n-1}^{*}$ for $\{e_{0}, \ldots, e_{n-1}\}$ an oriented orthonormal basis, although it has square-norm $-1$ (to correct this we should multiply it by $i$, but we are on a real vector space). We use the convention $\norm{e_{0}}^{2} = -1$. In this case, there is sometimes, but not always, a sign change with respect to the Euclidean case. For example, $*(e_{1}^{*} \wedge \ldots \wedge e_{n-1}^{*}) = (-1)^{n-1}e_{0}^{*}$, exactly as in the Euclidean case, since \eqref{HodgeStar} becomes, for $\alpha = \beta = (e_{1}^{*} \wedge \ldots \wedge e_{n-1}^{*})$:
	\[(e_{1}^{*} \wedge \ldots \wedge e_{n-1}^{*}) \wedge (-1)^{n-1}e_{0}^{*} = \langle (e_{1}^{*} \wedge \ldots \wedge e_{n-1}^{*}), (e_{1}^{*} \wedge \ldots \wedge e_{n-1}^{*}) \rangle \cdot \vol
\]
which is true since both the l.h.s. and the r.h.s. are equal to the volume form. Instead, $*(e_{0}^{*}) = -e_{1}^{*} \wedge \ldots \wedge e_{n-1}^{*}$, while in the Euclidean case there is no minus sign. In fact, \eqref{HodgeStar} becomes, for $\alpha = \beta = e_{0}^{*}$:
	\[e_{0}^{*} \wedge (-e_{1}^{*} \wedge \ldots \wedge e_{n-1}^{*}) = \langle e_{0}^{*}, e_{0}^{*} \rangle \vol
\]
and this is true because $\langle e_{0}^{*}, e_{0}^{*} \rangle = -1$ in the minkoskian case. Thus, given a summand $x \cdot e_{i_{1}}^{*} \wedge \ldots \wedge e_{i_{p}}^{*}$, its Hodge duals in the Euclidean and Minkowskian cases are equal if $0$ is not one of the indices $i_{1}, \ldots, i_{p}$, and they are opposite otherwise. In particular, since applying $*^{2}$ to any summand of this form we get the index $0$ one of the two times, it follows that $*^{2}$ in the Minkowskian and Euclidean cases are always opposite (we recall that in the eulidean case $*^{2}\vert_{\Lambda^{p}V^{*}} = (-1)^{p(n-p)}$, as it is easy to verify).

\section{Direct sum and direct product}\label{DirectSumProd}

We consider abelian groups, but all the discussion applies equally to the case of rings, vector spaces, or in general objects of a fixed abelian category. Given a family of abelian groups $\{G_{\alpha}\}_{\alpha \in I}$, we define the \emph{direct sum}:
	\[\bigoplus_{\alpha \in I} G_{\alpha}
\]
as the group whose elements are families made by one element for each group $G_{\alpha}$, such that only finitely many of them are non-zero; the sum is defined componentwise. Thus, an element of $G$ is a collection $\{g_{\alpha}\}_{\alpha \in I}$ for $g_{\alpha} \in G_{\alpha} \,\forall \alpha \in I$ and such that there exists a \emph{finite} set $J \subset I$ such that $g_{\alpha} = 0 \, \forall \alpha \in I \setminus J$. Instead, we define the \emph{direct product}:
	\[\prod_{\alpha \in I} G_{\alpha}
\]
as the group whose elements are families made by one element for each group $G_{\alpha}$, without any restriction. The direct sum is naturally a subgroup of the direct product; when the family is finite they coincide (in particular, the direct sum and the direct product of two groups coincide).

For $G^{*} := \Hom(G, \mathbb{Z})$ the following realtions hold:
	\[\Bigl(\bigoplus_{\alpha \in I} G_{\alpha}\Bigr)^{*} = \prod_{\alpha \in I} G_{\alpha}^{*} \qquad\qquad \Bigl(\prod_{\alpha \in I} G_{\alpha}\Bigr)^{*} \supset \bigoplus_{\alpha \in I} G_{\alpha}^{*} \; .
\]
In fact, in order to give a homomorphism $\varphi$ from $\bigoplus_{\alpha \in I} G_{\alpha}$ to $\mathbb{Z}$ it is enough to specify its restriction on each single group $G_{\alpha}$, since such groups generates their direct sum; thus, the homomorphism $\varphi$ is specified by a collection $\{\varphi_{\alpha}\}_{\alpha \in I}$ for $\varphi_{\alpha} \in G_{\alpha}^{*} \,\forall \alpha \in I$. We do not have to impose a finitness condition, since, even if there are infinitely many non-zero homomorphisms in the family, when we apply them to an element of the direct sum they can assume a non-zero value only on the non-zero elements, which are a finite set. That's why every element of $\prod_{\alpha \in I} G_{\alpha}^{*}$ gives a well-defined homomorphism from $\bigoplus_{\alpha \in I} G_{\alpha}$ to $\mathbb{Z}$. Instead, for the direct product, given a family $\{\varphi_{\alpha}\}_{\alpha \in I}$ for $\varphi_{\alpha} \in G_{\alpha}^{*} \,\forall \alpha \in I$, it gives a well-defined homomorphism from $\prod_{\alpha \in I} G_{\alpha}$ to $\mathbb{Z}$ if and only $\varphi_{\alpha} \neq 0$ only for finitely many elements. In fact, let us define $J \subset I$ as the set such that $\varphi_{\alpha} \neq 0$ if and only if $\alpha \in J$ and let us suppose that $J$ is infinite. Then, for each $\alpha \in J$, there exists $g_{\alpha} \in G_{\alpha}$ such $\varphi_{\alpha}(g_{\alpha}) = n_{\alpha} > 0$. If we choose any element $g_{\alpha}$ for $\alpha \in I \setminus J$, we obtain that $\{\varphi_{\alpha}\}(\{g_{\alpha}\})$ is an infinite sum, thus it is not well-defined. That's why only the elements of $\bigoplus_{\alpha \in I} G_{\alpha}^{*}$ give a well-defined homomorphism from $\prod_{\alpha \in I} G_{\alpha}$ to $\mathbb{Z}$. In this case we have just an inclusion, since it is not true that the single groups $G_{\alpha}$ generate the direct product: actually, the subgroup of the direct product generated by the single groups is exactly the direct sum, since in a group we allow only finite sums.


\end{document}